\documentclass[acmsmall]{acmart}
\AtBeginDocument{%
  \providecommand\BibTeX{{%
    \normalfont B\kern-0.5em{\scshape i\kern-0.25em b}\kern-0.8em\TeX}}}

\usepackage{algorithm}
\usepackage{algorithmic}
\usepackage{graphicx}
\usepackage{textcomp}
\usepackage{xcolor}
\usepackage{multirow,color, hyperref, url, graphics}
\usepackage{listings}
\usepackage{diagbox}
\usepackage{subfig}
\usepackage{balance}
\usepackage{adjustbox}
\usepackage{colortbl}
\usepackage{framed}
\usepackage{booktabs}
\usepackage{indentfirst}
\usepackage{makecell,tcolorbox}
\usepackage[framemethod=TikZ]{mdframed}
\usepackage{placeins}

\usepackage{array}

\def\renderinsertedpart{}

\ifx\renderinsertedpart\undefined
\newcommand{\insertedpart}[1]{#1}
\newenvironment{insertedpartenv}{}{}
\else
\newcommand{\insertedpart}[1]{#1}

\fi

\newcommand{\finding}[1]{
  \vspace{1.5mm}
 \begin{mdframed}[linecolor=gray,roundcorner=12pt,backgroundcolor=gray!15,linewidth=3pt,innerleftmargin=2pt, leftmargin=0cm,rightmargin=0cm,topline=false,bottomline=false,rightline = false]
  #1
 \end{mdframed}
 \vspace{1.5mm}
}

\begin{document}

\title{LogiDroid: Individual Functional Test Generation via Business Logic Extraction and Adaptation}

\author{Yakun Zhang}
\email{zhangyk@hit.edu.cn}
\affiliation{%
  \department{Shenzhen Key Laboratory of Internet Information Collaboration}
  \institution{Harbin Institute of Technology, Shenzhen}
  \city{Shenzhen}
  \country{China}
}

\author{Zihan Wang}
\email{2023112983@stu.hit.edu.cn}
\affiliation{%
  \institution{Harbin Institute of Technology}
  \city{Harbin}
  \country{China}
}

\author{Xinzhi Peng}
\email{2023311107@stu.hit.edu.cn}
\affiliation{%
  \department{Shenzhen Key Laboratory of Internet Information Collaboration}
  \institution{Harbin Institute of Technology, Shenzhen}
  \city{Shenzhen}
  \country{China}
}

\author{Zihao Xie}
\email{25s151153@stu.hit.edu.cn}
\affiliation{%
  \department{Shenzhen Key Laboratory of Internet Information Collaboration}
  \institution{Harbin Institute of Technology, Shenzhen}
  \city{Shenzhen}
  \country{China}
}

\author{Xiaodong Wang}
\email{wang_xiaodong@stu.pku.edu.cn}
\affiliation{%
  \institution{Key Lab of HCST (PKU), MOE; SCS, Peking University}
  \city{Beijing}
  \country{China}
}

\author{Xutao Li}
\email{lixutao@hit.edu.cn}
\affiliation{%
  \department{Shenzhen Key Laboratory of Internet Information Collaboration}
  \institution{Harbin Institute of Technology, Shenzhen}
  \city{Shenzhen}
  \country{China}
}

\author{Dan Hao}
\email{haodan@pku.edu.cn}
\affiliation{%
  \institution{Key Lab of HCST (PKU), MOE; SCS, Peking University}
  \city{Beijing}
  \country{China}
}

\author{Lu Zhang}
\email{zhanglucs@pku.edu.cn}
\affiliation{%
  \institution{Key Lab of HCST (PKU), MOE; SCS, Peking University}
  \city{Beijing}
  \country{China}
}

\author{Yunming Ye}
\authornote{Corresponding author.}
\email{yym@hit.edu.cn}
\affiliation{%
  \department{Shenzhen Key Laboratory of Internet Information Collaboration}
  \institution{Harbin Institute of Technology, Shenzhen}
  \city{Shenzhen}
  \country{China}
}

\renewcommand{\shortauthors}{Zhang et al.}

\begin{abstract}

Functional testing is essential for verifying that the business logic of mobile applications aligns with user requirements, serving as the primary methodology for quality assurance in software development. Despite its importance, functional testing remains heavily dependent on manual effort due to two core challenges. First, acquiring and reusing business logic from unstructured requirements remains difficult, which hinders the understanding of specific functionalities. Second, a significant semantic gap exists when adapting business logic to the diverse GUI environments, which hinders the generation of test cases for specific mobile applications.

To address the preceding challenges, we propose \emph{\toolNameSmall{}}, a two-stage approach that generates individual functional test cases by extracting business logic and adapting it to target applications.
First, in the Knowledge Retrieval and Fusion stage, two LLM-based agents (i.e., a Semantic-Retrieval Agent and a Knowledge-Fusion Agent), are employed to construct a functional test dataset, retrieve relevant test cases, and extract structured business logic for the target functionality.
Second, in the Context-Aware Test Generation stage, two LLM-based agents (i.e., a Perception–Interaction Agent and a Decision–Generation Agent), jointly analyze the extracted business logic and the real time GUI environment to incrementally generate context adaptive functional test cases.
This design allows \toolNameSmall{} to accurately understand application semantics and use domain expertise to generate complete test cases with verification assertions.
We assess the effectiveness of \toolNameSmall{} using two widely-used datasets that cover 28 real-world applications and 190 functional requirements. Experimental results show that \toolNameSmall{} successfully tested 40\% of functional requirements on the FrUITeR dataset
(an improvement of over 25\% compared to the state-of-the-art approaches)
and 65\% on the Lin dataset (an improvement of over 55\% compared to the state-of-the-art approaches). These results demonstrate the significant effectiveness of \toolNameSmall{} in functional test generation.

\end{abstract}


\acmJournal{TOSEM}
\acmVolume{37}
\acmNumber{4}
\acmArticle{111}
\acmMonth{12}
\acmDOI{10.1145/XXXXXXX.XXXXXXX}


\graphicspath{{figures/}}
\newcommand{\figref}[1]{Figure~\ref{#1}}
\newcommand{\tabref}[1]{Table~\ref{#1}}
\newcommand{\secref}[1]{Section~\ref{#1}}

\newcommand{\toolName}[1]{\textbf{MACdroid\raisebox{-0.7ex}{\Huge\textbf{#1}}}}

\newcommand{\toolNameplain}[1]{MACdroid\raisebox{-0.7ex}{\Huge #1}}

\newcommand{\toolNameSmall}{LogiDroid}

\makeatletter
\let\@ACM@checkaffil\relax
\makeatother

\maketitle
\newcommand{\yakun}[1]{\textcolor{blue}{[yakun: #1]}}

\newcommand{\ie}{\mbox{\emph{i. e.,\ }}}

\section{Introduction}\label{sec:introduction}
In the past decade, mobile applications have evolved into an indispensable form of software. As of 2025, over four million applications are available in application stores \cite{app-statistics}, which makes quality assurance a critical priority for developers. Among various quality assurance methodologies, functional testing is essential as it directly verifies the business logic and user experience from an end-user perspective.

Functional testing for mobile applications primarily relies on generating \emph{individual test cases} for different functionalities accessed through Graphical User Interfaces (GUIs). A GUI test case typically involves executing an ordered sequence of \emph{events} on GUI \emph{widgets} that belong to specific GUI \emph{states}, accompanied by \emph{assertions} to verify whether the outcomes align with developer expectations. However, existing automated testing approaches~\cite{su2017guided,baek2016automated,gu2017aimdroid,lai2019goal, li2017droidbot, su2021fully, wang2022detecting, liu2022guided, yu2024practical} focus mainly on defect detection, such as identifying crashes or resource leaks. These approaches often lack a deep semantic understanding of application functionalities, rendering them unable to generate test cases targeted at specific business logic.
With the emergence of Large Language Models~\cite{achiam2023gpt,singh2025openai,chen2025deep} (LLMs), several recent approaches~\cite{zhang2025appagent, wang2025llmdroid, wen2023empowering, li2025reusedroid, zhang2026gui} explore the target applications and leverage the reasoning capabilities of LLMs to generate GUI event sequences. However, these approaches rely on general-purpose knowledge, which is insufficient for understanding application-specific functionalities and adapting to the dynamic GUI behaviors required for complex functional verification. Moreover, they primarily generate event sequences and are unable to produce assertions, which limits their practical value for real-world functional testing.
As a result, functional testing remains heavily dependent on manual effort. This reliance not only limits efficiency and scalability but also introduces quality risks due to subjective inconsistencies.

Despite the increasing demand for functional testing automation, achieving high-quality functional test generation still faces two core challenges.

\underline{Challenge 1: Acquisition and Reuse of Business Logic.} The generation of high-quality functional test cases  relies heavily on understanding the complex business logic of target functionalities~\cite{kamimura2015measuring,metin2025business, zhang2024learning, zhang2024synthesis}. This logic encompasses the specific operational rules and decision-making processes that define a functional requirement. For example, the business logic of a ``registration'' functionality including entering the registration state, filling all the necessary inputs, and validating the registration result through specific state changes. However, such expertise typically resides in unstructured natural language descriptions or scattered historical test cases, making it difficult to formalize~\cite{arora2024generating,liu2024exploring}.
Meanwhile, since recent approaches still mainly rely on general knowledge from the foundation models, they lack an explicit mechanism to extract, represent, and reuse business logic across applications, making them difficult to generate effective functional test cases.

\underline{Challenge 2: Semantic understanding and adaptation of GUI environments.} Even when business logic is available, grounding it in the diverse and dynamic GUI environments of mobile applications remains a significant challenge. Each application implements its functionalities through unique GUI widgets, intricate interaction flows, and application-specific layout structures~\cite{zhang2024learning}. For example, a ``registration'' functionality may involve entirely different widgets, state transitions, and validation assertions across different shopping applications. As a result, the same functionality-level logic may correspond to different concrete GUI operations in different applications. This semantic gap makes it difficult to map abstract business logic to the specific GUI implementation of a target application~\cite{zhang2026gui}, often leading to generated test cases that fail to execute the intended business scenario accurately. Such difficulty is further increased by existing approaches that mainly treat functional test generation as direct exploration on the target application, without an explicit mechanism to bridge functionality-level knowledge and GUI-level execution.

We observe that although the functionalities of mobile applications vary, those with similar functionalities often exhibit related business logic and testing patterns. This observation indicates that knowledge reuse can be leveraged to generate individual test cases effectively.
To this end, we propose \textbf{\toolNameSmall{}}, a two-stage approach that generates individual functional test cases through business logic extraction and adaptation to the target application.
Detailed two stages are as follows.

\underline{Stage 1: Knowledge Retrieval and Fusion.} This stage involves two LLM-based agents: a \emph{Semantic-Retrieval Agent} and a  \emph{Knowledge-Fusion Agent}. The Semantic-Retrieval Agent is responsible for constructing a functional test dataset (containing 294 functional test cases from 71 applications) and automatically retrieving relevant test cases related to the target functionality within the dataset. Furthermore, the Knowledge-Fusion Agent semantically aligns and fuses the retrieved test cases, and extracts the business logic for the target functionality.
\textbf{This stage establishes a transformation channel from vague requirements to structured business logic for subsequent test generation, effectively addressing the \emph{Challenge 1}.}

\underline{Stage 2: Context-Aware Test Generation.} This stage relies on the close collaboration between two LLM-based agents: a \emph{Perception–Interaction Agent} and a \emph{Decision–Generation Agent}. The Perception–Interaction Agent serves as the interaction interface between \toolNameSmall{} and the target application. It continuously captures multimodal data of target application (e.g., state images and widget texts), thereby providing rich context for decision-making. The Decision–Generation Agent acts as the core reasoning engine of \toolNameSmall{}. By jointly analyzing the business logic provided in Stage 1 and the real-time contextual information, it decomposes the testing task into multiple steps and incrementally generates context-adaptive test cases.
\textbf{This stage maintains the depth of knowledge guidance during test generation while enhancing dynamic adaptability to environmental changes, effectively addressing the \emph{Challenge 2}.}

We conduct a comprehensive evaluation to analyze the effectiveness of \toolNameSmall{} using 28 real-world mobile applications, 190 functional requirements, and corresponding test cases from two popular datasets (i.e., the FrUITeR dataset~\cite{fruiter-dataset} and the Lin dataset~\cite{craftdroid-dataset}). We compare \toolNameSmall{} with the state-of-the-art (sota) approaches from both academia and industry, i.e., AutoDroid~\cite{wen2023empowering} and AppAgent~\cite{zhang2025appagent}.
On the FrUITeR dataset, the test cases generated by \toolNameSmall{} successfully validate 40\% of the target functionalities, representing a 25\% improvement over the baselines. On the Lin dataset, \toolNameSmall{} successfully tests 65\% of the target functionalities, outperforming the baselines by 55\%. \emph{Note that, \toolNameSmall{} is able to generate complete test cases with assertions that the baselines fail to produce.} Overall, these results demonstrate that \toolNameSmall{} is effective in functional test generation for industrial applications.

The main contributions of this research are summarized as follows:
\begin{itemize}
    \item \textbf{Methodological innovation.}  Given the requirement description of a specific functionality, we propose LogiDroid, a novel approach that generates individual functional test cases (with assertions), which makes the large-scale functional testing possible in industry.

    \item \textbf{Technical design.} We design (1) a novel Knowledge Retrieval and Fusion technique to provide a reliable business logic for test generation; and (2) a novel Context-Aware Test Generation technique that combines multimodal information with domain knowledge to incrementally generate test cases adapted to the target application.

    \item \textbf{System implementation.} We develop a complete prototype tool and construct a functional test dataset containing 294 functional test cases from 71 applications across 13 categories. To promote research reproducibility, \textbf{the source code has been released as open source~\cite{logidroid}}.

    \item \textbf{Experimental evaluation.} We conduct an empirical evaluation using real-world applications, demonstrating the effectiveness of \toolNameSmall{}.
\end{itemize}

\begin{figure*}[t]
	\center
 \includegraphics[width=1\linewidth]{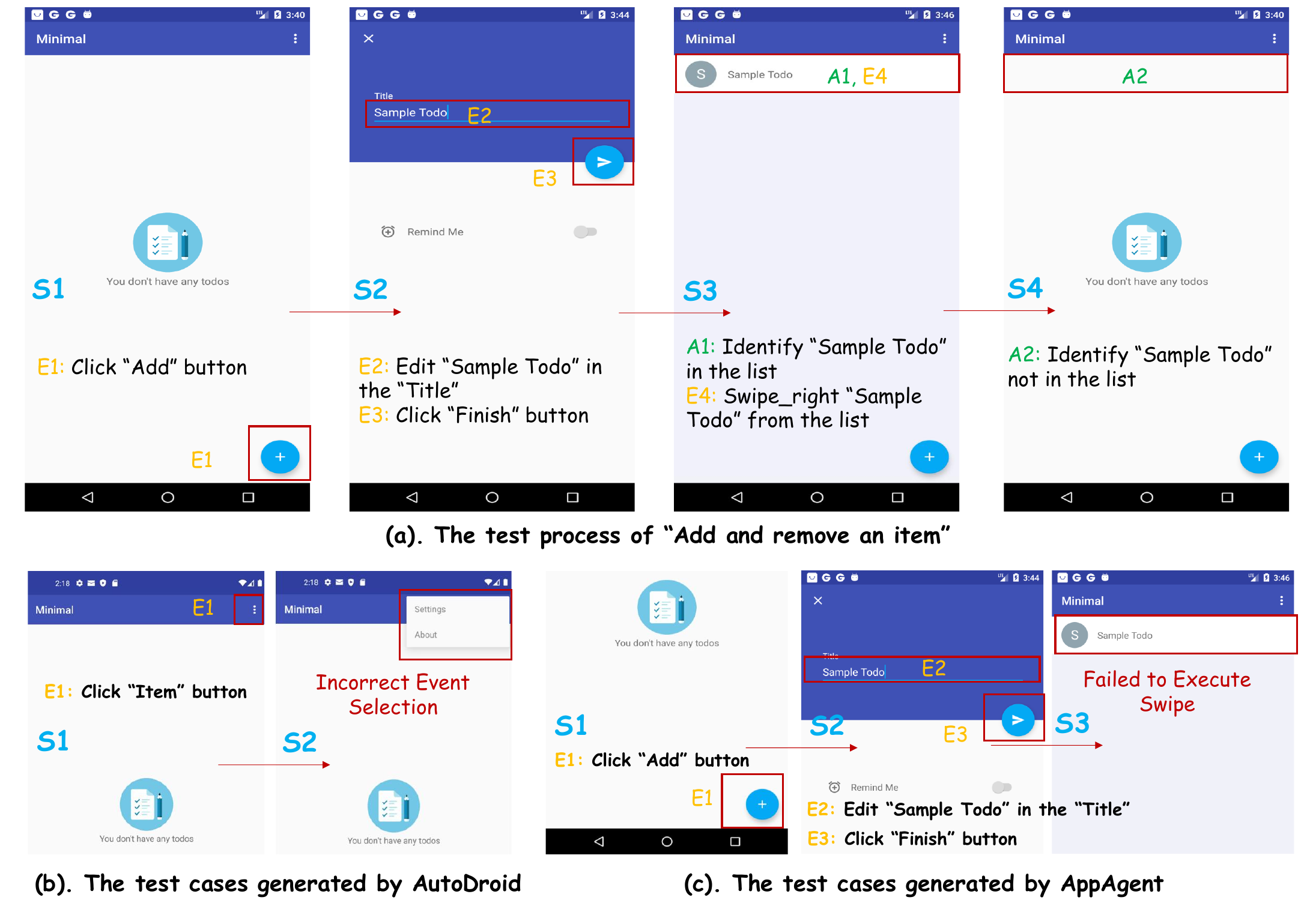}
	\caption{The test process of ``Add and remove an item'' and the corresponding results of AutoDroid and AppAgent}
	\label{fig:illustrated-example}
\end{figure*}

\section{Preliminaries}

\subsection{GUI Test Cases}

A GUI test case (e.g., \figref{fig:illustrated-example}(a)) typically consists of an ordered sequence of \emph{events} performed on \emph{widgets} across \emph{states}, together with \emph{assertions} that validate the observed outcomes against expected results. A GUI \emph{state} (e.g., S1 in \figref{fig:illustrated-example}(a)) denotes the observable interface of an application at a particular moment during execution, including the current screen and the visible GUI widgets on it. A GUI \emph{widget} (e.g., ``Add'' button in  \figref{fig:illustrated-example}(a)) denotes an interactive element, such as a button, an input field, or a list item. An \emph{event} (e.g., E1 in \figref{fig:illustrated-example}(a)) denotes an action performed on a widget, such as clicking, editing, or swiping. An \emph{assertion} (e.g., A1 in \figref{fig:illustrated-example}(a)) denotes a checking step that determines whether an expected condition holds in the current or a subsequent GUI state, such as whether a target widget appears or disappears.

Accordingly, a GUI test case is a sequence of state transitions driven by events and validated by assertions, as illustrated in \figref{fig:illustrated-example}. This terminology serves as the basis of \toolNameSmall{}, which takes a functional requirement as input and generates a functional test case for the target application.

\subsection{Illustrative Example}
\figref{fig:illustrated-example} illustrates the test workflow of the ``Add and remove an item'' functionality in a to-do application~\cite{craftdroid-dataset}. The test case is designed to validate whether a user can successfully add a to-do item and subsequently remove it after finishing the task. We use this case to explain our design motivation. It also helps to illustrate the working mechanism of \toolNameSmall{} in later sections.

The test case comprises four GUI states (S1-S4), four events (E1-E4), and two assertions (A1-A2). The test process begins by triggering a click event on the Add button (E1). Subsequently, the text ``Sample todo'' is input into the Title box (E2), followed by a click operation on the Finish button (E3). Assertion A1 confirms the successful addition of an item by checking whether the text ``Sample todo'' appears in a new state (S3). Then, a swipe-right operation (E4) is performed on the added item. Finally, assertion A2 validates the removal of the item by detecting whether the item disappears from the state (S4).

We evaluate the performance of two representative functional testing approaches, AutoDroid~\cite{wen2023empowering} and AppAgent~\cite{zhang2025appagent}, on the aforementioned test functionality. As shown in \figref{fig:illustrated-example}(b) and \figref{fig:illustrated-example}(c), AutoDroid and AppAgent both fail to generate valid test cases for this functionality. This limitation primarily arises from the lack of effective mechanisms for domain knowledge acquisition and reuse. Consequently, they struggle to understand the functional semantics deeply and generate useful event sequences. Specifically, AutoDroid deviates from the intended workflow by selecting an incorrect event, while AppAgent fails to execute the required swipe interaction and therefore cannot complete the item removal functionality. Furthermore, they fail to generate the necessary verification assertions for the target application.

\textbf{The significant disparity between test cases generated by existing approaches and target functionalities hinders their direct application in industrial scenarios.} For this reason, functional testing for the mobile application still relies on extensive manual correction and maintenance. Therefore, it is imperative to develop innovative approaches capable of generating high-quality functional test cases.

\section{\toolNameSmall}\label{sec:ragdroid}

\begin{figure}[t]
	\center
 \includegraphics[width=1\linewidth]{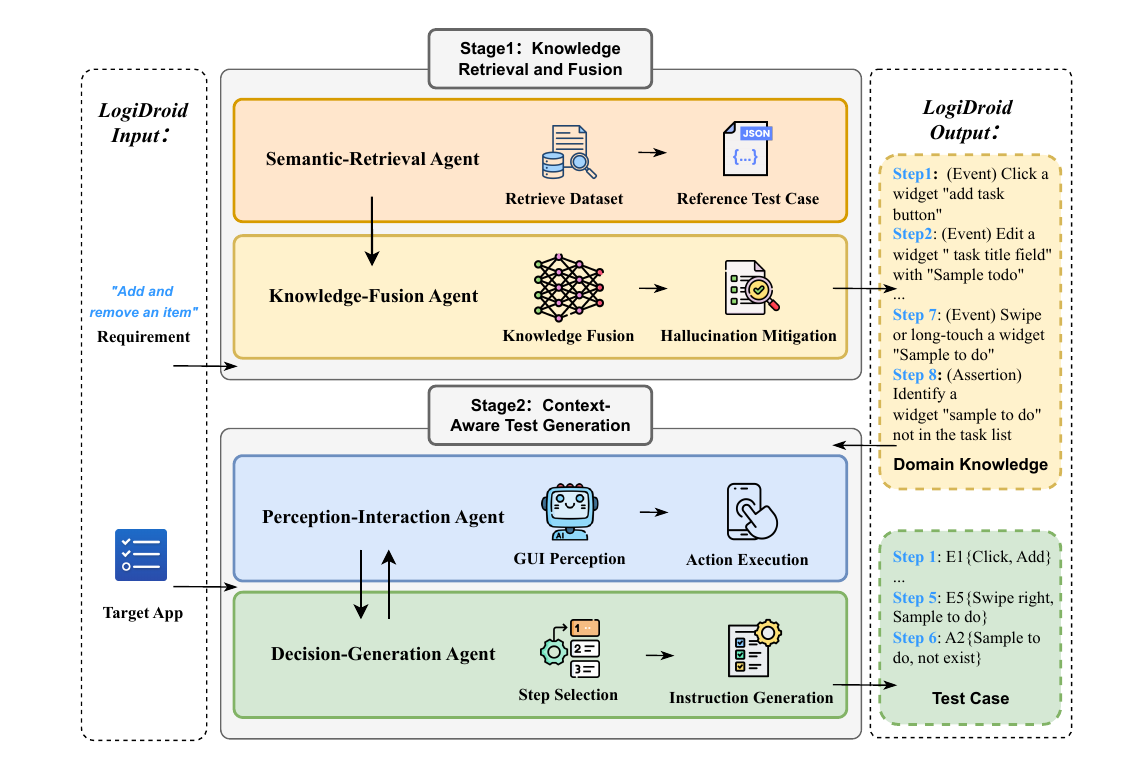}
	\caption{Overview of \toolNameSmall{}}
	\label{fig:overview}
\end{figure}

Given the target application and its functional requirements, \toolNameSmall{} automatically generates test cases to verify the corresponding functionalities. As illustrated in \figref{fig:overview}, the overall workflow of \toolNameSmall{} consists of two primary stages.
The first stage is the \textbf{Knowledge Retrieval and Fusion}. This stage aims to retrieve test cases from historical repositories that are semantically relevant to the current functional requirements. It then distills business logic for the corresponding functionalities, providing a reliable domain knowledge to guide subsequent test generation (see \secref{sec:retrieval}).
The second stage is the \textbf{Context-Aware Test Generation}. In this stage, \toolNameSmall{} explores the target application in real-time to capture multimodal information, including GUI layouts and visual screenshots. By integrating the business logic acquired from the first stage with this multimodal context, the system employs a progressive decision-making mechanism to decompose functional requirements into executable test sequences. Note that, unlike existing approaches~\cite{zhang2025appagent,wen2023empowering,li2017droidbot} that only generate events, \toolNameSmall{} generates comprehensive test cases with both events and  assertions (see \secref{sec:generation}).

\subsection{Knowledge Retrieval and Fusion}
\label{sec:retrieval}

Given the specific functionality to be verified, the Knowledge Retrieval and Fusion stage aims to retrieve functionally similar instances from existing real-world test cases and distill expert-level domain knowledge from them. However, accurately matching the functional requirements within a massive repository of test cases and subsequently extracting reusable domain knowledge remains \textbf{\emph{a critical challenge}}. To address this challenge, \toolNameSmall{} utilizes a dual-agent configuration consisting of the Semantic-Retrieval Agent and the Knowledge-Fusion Agent, as depicted in \figref{fig:overview}.

Specifically, we construct a functional test dataset including functional test cases for diverse functionalities. The \textbf{Semantic-Retrieval Agent} retrieves relevant test cases within the dataset (see \secref{sec:search}). Subsequently, the \textbf{Knowledge-Fusion Agent} processes these retrieved cases to distill domain knowledge into structured documentation. This documentation captures the business logic and expert-level testing patterns required to detect the target functionality, serving as a foundation for guiding the subsequent generation process (see \secref{sec:align}).

\subsubsection{Semantic-Retrieval Agent}
\label{sec:search}

In industrial software environments, large repositories of functional test cases contain domain expertise and business logic essential for validating specific functionalities. Despite their significant reference value, these test cases are frequently scattered across heterogeneous sources, which makes their effective use difficult. Therefore, it is crucial to  construct a high-quality test dataset and develop mechanisms to retrieve the relevant test cases.

To address the preceding challenge, \toolNameSmall{} automatically constructs the functional test dataset and designs the test case retrieval. This process is divided into two main components: \emph{Dataset Construction} and \emph{Test Case Retrieval}.

\textbf{Dataset construction.} This component details the systematic collection of functional test cases and the methodology for generating functional summaries.

\underline{Test case collection.} To construct the functional test dataset, we systematically collect test cases through the following two channels.

(1) Existing datasets. We collect 95 valid test cases from two established datasets in the application testing domain: the Lin dataset \cite{craftdroid-dataset} and the FrUITeR dataset \cite{fruiter-dataset}. These cases from 28 applications cover six common application scenarios, such as news and shopping.

(2) Open-source projects. To further enhance the diversity and coverage, we systematically collect additional test cases from F-Droid \cite{fdroid}, a prominent open-source repository for mobile applications. Similar to related research~\cite{zhao2020fruiter, su2017guided, hu2018appflow}, we focus on eight popular categories (e.g., navigation, sport, and device) including one common category (i.e., shopping) with existing datasets. We firstly select the top 30 applications by download volume in each category. 
We then exclude applications that do not contain  test cases, remaining 43 F-Droid applications with 199 functional test cases.

\underline{Functional summary generation.} To facilitate semantic retrieval, we generate a concise functional summary for each test case to encapsulate the core functionalities it verifies.

Given the advanced comprehension and summarization capabilities of LLMs, we implement an LLM-based automated generation approach as an alternative to traditional manual authorship or rule-based techniques. This approach offers superior scalability for future dataset expansions while maintaining consistency in quality and semantic accuracy. To guide the LLM in accurately comprehending and executing the summarization task, we design a structured prompt template. This design is motivated by the fact that LLMs demonstrate superior performance when processing structured information, as such formats align more closely with the organizational patterns found in their underlying training data~\cite{liu2023chatting, zhang2026gui, feng2024prompting}. As illustrated in \tabref{tab:prompt_pattern}, the template consists of the following four parts.

\newcommand{\vc}[2]{\raisebox{-#1ex}{#2}}

\begin{table}[t]
\centering
\caption{Prompt Template for Functional Summary Generation}

\begin{tabular}{l|p{10cm}}
\toprule
\textbf{AIM} & \textbf{EXAMPLE} \\
\midrule

\vc{1.6}{Task Definition} &
You are a functional summary generator. Based on the test cases for the Android app,
generate a natural and one-sentence description. \\

\midrule
\vc{9.5}{Input Object} &
Test case from a \textbf{[To-do]} app\newline
Step 1: (Event) Click a widget ``add todo item button'' \newline
Step 2: (Event) Edit a widget ``user todo edit text'' with ``sample todo'' \newline
Step 3: (Event) Click a widget ``make todo floating action button'' \newline
Step 4: (Assertion) Identify a widget ``sample to do'' in the state \newline
Step 5: (Event) Swipe right a widget ``sample to do'' \newline
Step 6: (Assertion) Identify a widget ``sample to do'' not in the state \newline
Functional summary:\\

\midrule
\vc{7}{Demonstration Case} &
\textbf{Example 1}: Test case from a Browser app \newline
Step 1: (Event) Click a widget ``search'' \newline
Step 2: (Event) Edit a widget ``search'' with ``news'' \newline
Step 3: (Event) Identify a widget ``latest news'' in the state \newline
Functional summary: Test the search functionality\newline
\textbf{Example 2}:...\\

\midrule
\vc{5.5}{Acceptance Criteria} &
Please generate the functional description for the \textbf{[To-do]} app.\newline
1. Please keep it simple: only include at most the subject, verb, and object. \newline
2. Please use natural English, not technical terms. \newline
3. Please focus on the main actions, ignore the details. \\

\bottomrule
\end{tabular}
\label{tab:prompt_pattern}
\end{table}

(1) Task Definition. This part explicitly defines the core objective for the LLM, which is to distill the business logic of a complex functional test case into a concise and one-sentence summary. By establishing this specific goal, the model focuses on high-level functional intent rather than low-level implementation details.

(2) Input Object. This part provides the test case to be summarized alongside its corresponding application category. To facilitate model comprehension, \toolNameSmall{} imports data in JSON format, decomposing each test case into a series of events and assertions in a semi-structured representation. Specifically, each \textbf{event} includes a widget identifier, an operation type, and an optional parameter (as shown in ``Step 2'' of \tabref{tab:prompt_pattern}). Each \textbf{assertion} includes a widget identifier and a verification condition (as shown in ``Step 4'' of \tabref{tab:prompt_pattern}). To achieve both precision and conciseness, each widget is characterized solely by three key attributes: \texttt{text}, \texttt{resource-id}, and \texttt{content-desc}.

(3) Demonstration Case. Leveraging the powerful in-context learning capabilities of LLMs~\cite{dong2022survey}, we provide two complete examples within the prompt template. These examples clearly demonstrate the mapping between raw test cases and their functional summaries, enabling the LLM to align its output with the required format and logic.

(4) Acceptance Criteria. To constrain the model's stochastic behavior and minimize the risk of hallucinations~\cite{ji2023towards}, this part enforces three constraints: (i) the output must be concise, adhering to a basic subject-verb-object structure; (ii) the summary must be in natural language rather than code; and (iii) the summary must focus on the primary operations of the test case while disregarding secondary implementation details. These criteria collectively ensure the precision and readability of the generated content.

\begin{table}[t]
  \centering
  \caption{The statistics of Functional Test dataset of \toolNameSmall{}}

    \begin{tabular}{l|r|r|r|r}
    \toprule
    \multicolumn{1}{c|}{\textbf{Category}} & \multicolumn{1}{c|}{\textbf{Application}} & \multicolumn{1}{c|}{\textbf{Test}} &
    \multicolumn{1}{c|}{\textbf{Summary}} &
    \multicolumn{1}{c}{\textbf{Functionality}}
    \\
    \midrule
    News  & 4  & 32 & 32 & 12\\
    Shopping & 7  & 29 & 29 & 12\\
    Browser & 12 & 51 & 51 & 10\\
    To-do & 5  & 10 & 10 & 2\\
    Mail  & 2  & 4  & 4  & 2\\
    Calculator & 5  & 10 & 10 & 2\\
    Note  & 6  & 28 & 28 & 14\\
    Navigation & 3  & 6  & 6  & 4\\
    Draw  & 3  & 9  & 9  & 5\\
    System & 9  & 46 & 46 & 11\\
    Device & 7  & 29 & 29 & 9\\
    Sport & 3  & 24 & 24 & 11\\
    Time  & 5  & 16 & 16 & 5\\
    \midrule
    \textbf{Total} & \textbf{71} & \textbf{294} & \textbf{294} & \textbf{99}\\
    \bottomrule
    \end{tabular}%
  \label{tab:database}%
\end{table}%

\underline{Dataset Statistics.} The final dataset includes 13 common categories, 99 functionalities, across 71 mobile applications, with 294 functional test cases and their corresponding summaries. The statistical details of the dataset are summarized in \tabref{tab:database}. To support efficient retrieval, \toolNameSmall{} implements a structured storage format where the application category serves as the primary index ($Key\_1$), the embedding vector of the functional summary serves as the secondary index ($Key\_2$), and the corresponding test case is stored as the value ($Value$). By integrating category-based filtering with vector-based semantic matching, this multi-dimensional indexing mechanism enables the precise and efficient identification of specific functionalities.

\textbf{Test case retrieval.}
Regarding the test requirements provided by the user, \toolNameSmall{} first encodes the requirements into an embedding vector and subsequently performs a similarity-based retrieval within the functional test dataset. By calculating the cosine similarity between the user's requirement vector and the functional summary vectors, \toolNameSmall{} selects the $Top_{sim}$ results that exhibit the highest similarity scores and belong to the same application category.
The test cases associated with these results contain valuable business logic, which serves as a critical reference for the subsequent test generation process and providing the necessary knowledge foundation for verifying specific functionalities.

\subsubsection{Knowledge-Fusion Agent}
\label{sec:align}

To effectively reuse the domain expertise, the Knowledge-Fusion Agent focuses on extracting \emph{core business logic} from a set of relevant test cases provided by the semantic-retrieval agent. For instance, in the context of registration functionalities, while specific implementations vary across different mobile applications (e.g., utilizing email-password combinations or username-phone number verification), the core business logic follows a standardized pattern. This pattern typically involves navigating to the registration state, inputting valid information, executing the registration action, and verifying the resultant state.

The primary objective of the Knowledge-Fusion Agent is to extract implementation-agnostic core business logic from multiple relevant test cases. By extracting away concrete implementation details, this agent provides reusable domain knowledge for subsequent test generation, effectively emulating the strategic guidance of a human testing expert.

\textbf{Knowledge fusion.} The Knowledge-Fusion Agent provides a structured fusion framework based on LLMs, with the template architecture illustrated in \tabref{tab:prompt_fusion}. This framework comprises four key parts designed to facilitate the extraction of reusable testing patterns.

First, in the ``Task Definition'', the model is explicitly directed to extract generalized business logic by fusing information from multiple retrieved test cases and the target requirement. Second, the ``Input Object'' supplies the set of relevant test cases identified by the Semantic-Retrieval Agent. Third, the ``Demonstrate Case'' provides a complete demonstration of the abstraction and fusion process, clearly defining the expected output format. Finally, the ``Acceptance Criteria'' enforce three rigorous constraints to enhance the quality of the generated output, including (i) the number of logic steps must remain within a reasonable range to maintain efficiency; (ii) the generated steps must strictly adhere to specified formatting conventions, where events follow the format ``[Action] a widget [Widget] with [Value]'' and assertions follow the format``Identify a widget [Widget] [Condition]''; (iii) the response must present the core business logic directly without any auxiliary explanations or code instructions.

\begin{table}[t]
\centering
\caption{Prompt Template for Test Knowledge Fusion}

\begin{tabular}{l|p{10cm}}
\toprule
\textbf{AIM} & \textbf{EXAMPLE} \\
\midrule

\vc{1.5}{Task Definition} &
You are a summarizer to fuse test knowledge for the functionality: \textbf{[Add and remove an item]} in a \textbf{[To-Do]} app. \\

\midrule
\vc{7}{Input Object} &
\textbf{Related Test Case 1}: \newline
Step 1: (Event) Click a widget ``skip button'' \newline
Step 2: (Event) Click a widget ``new task button'' \newline
...\newline
Step 7: (Assertion) Identify a widget ``sample to do'' not in the state \newline
\textbf{Related Test Case 2}:... \\

\midrule
\vc{7.0}{Demonstrate Case} &
\textbf{Example}: Test knowledge for the functionality: [Test the search functionality] in a [Browser] app. \newline
Step 1: (Event) Click a widget ``search'' or ``url'' in the search bar \newline
Step 2: (Event) Edit a widget ``search'' or ``url'' in the search bar with ``news'' \newline
Step 3: (Event) Identify a widget ``latest news'' in the state \\

\midrule
\vc{7.0}{Acceptance Criteria} &
Please generate the test knowledge for the \textbf{[To-do]} app. \newline
1. The generated test step do not too short or too long. \newline
2. Please strictly use steps in the format of Event and Assertion \newline
(1) (Event) [Action] a widget [Widget] with [Value] \newline
(2) (Assertion) Identify a widget [Widget] [Condition] \newline
3. Please do not include any code, XPATH, or scripting instructions \\

\bottomrule

\end{tabular}
\vspace{-12pt}
\label{tab:prompt_fusion}
\end{table}

\textbf{Hallucination mitigation.} To address the challenge of hallucinations in LLMs~\cite{ji2023towards}, \toolNameSmall{} implements an robust automated detection and feedback mechanism. Through the application of predefined rules, this mechanism rigorously assesses whether the synthesized output satisfies the required formatting conventions, remains within the boundaries for testing steps, and meets the criteria for logical relevance.

When an invalid output is detected, the system automatically generates corrective feedback and triggers a re-generation process, which continues iteratively until the output satisfies all predefined specifications. This closed-loop verification process effectively safeguards the reliability of the knowledge fusion stage and ensures the high quality of the final output.

\subsection{Context-Aware Test Generation}
\label{sec:generation}

While the core business logic extracted from the first stage provides domain expertise for functional verification, significant implementation differences across various mobile applications prevent this logic from being directly applied to a target application. Bridging the gap between high-level business logic and application-specific events and assertions remains \emph{\textbf{a challenge}}. To address this challenge, we design two collaborative agents comprising the \emph{Perception-Interaction Agent} and the \emph{Decision-Generation Agent}, as illustrated in \figref{fig:overview}.

\begin{algorithm}[t]
\caption{Agent Interaction Loop (Stage 2)}
\label{alg:Agent Interaction}
\begin{algorithmic}[1]
\renewcommand{\algorithmicrequire}{\textbf{Input:}}
\renewcommand{\algorithmicensure}{\textbf{Output:}}

\REQUIRE App, Requirement $R$, Core Logic $L$
\ENSURE Test Case $T$

\WHILE{Task Not Complete}
    \STATE // Perception-Interaction Agent (PIA) \& Decision-Generation Agent (DGA)
    \STATE $S_{curr} \leftarrow PerceiveState(App)$ \hfill $\triangleright$ \textbf{PIA}: Capture multimodal state
    \STATE $Step \leftarrow StepSelection(R, L, S_{curr})$ \hfill $\triangleright$ \textbf{DGA}: Align logic with state
    \STATE $Instr \leftarrow InstructionGeneration(Step, S_{curr})$ \hfill $\triangleright$ \textbf{DGA}: Generate executable action
    \STATE $Execute(App, Instr)$ \hfill $\triangleright$ \textbf{PIA}: Interact with device
    \IF{$CompletionJudgment(Step, Instr, S_{curr})$}
        \STATE \hfill $\triangleright$ \textbf{DGA}: Verify step completion
        \STATE $Index \leftarrow Index + 1$
    \ENDIF
\ENDWHILE

\RETURN $T$
\end{algorithmic}
\end{algorithm}

Algorithm~\ref{alg:Agent Interaction} shows the interaction of between the two agents in the second stage.
First, the \textbf{Perception-Interaction Agent} (recalled Section \ref{sec:interaction}) is responsible for dynamically capturing GUI states of the target application, providing real-time environmental context for the Decision-Generation Agent (Line 3). Second, the \textbf{Decision-Generation Agent} (recalled Section \ref{sec:strategy}) synthesizes the multi-modal information of the target application with the core business logic from the first stage. This agent then generates an instruction sequence specifically adapted to the target application and its corresponding functionalities (Lines 4-5). Third, the \textbf{Perception-Interaction Agent} executing the instruction strategies derived from Decision-Generation Agent to the target application (Line 6). Through the alternating execution and closed-loop feedback of these two agents, this stage achieves an effective transformation from core business logic to executable test cases for specific functionalities.

\subsubsection{Perception-Interaction Agent}
\label{sec:interaction}

There are two primary objectives of the Perception-Interaction Agent. First, this agent explores the target application by dynamically retrieving multi-modal information, which encompasses both visual data from screenshots and textual data from the GUI hierarchy. This multi-modal information is subsequently sent to the Decision-Generation Agent to facilitate the generation of instruction sequences. Second, upon receiving the instruction sequences  produced by the Decision-Generation Agent, the Perception-Interaction Agent executes concrete operations on the target application. The Perception-Interaction Agent in a continuous loop operates continuously until a ``Task Complete'' signal is received, during which time it automatically records all interaction events and verification assertions to eventually synthesize them into a comprehensive functional test case.

The workflow of this agent consists of three critical components, which are GUI perception, action execution, and case synthesis.

\textbf{GUI Perception.} The Perception-Interaction Agent captures the visual and interactive widgets of the current state in real time by parsing the GUI hierarchy of the target application. This implementation involves two critical steps. First, this agent captures a screenshot of the current state to preserve complete visual information. Second, it transforms the state content into a structured natural language description. During this descriptive process, this agent extracts three core semantic attributes for each widget, specifically \texttt{text}, \texttt{content-desc}, and \texttt{resource-id}, while enumerating the operation types supported by each widget.

To maintain a logical structure for the preceding structured descriptions, \toolNameSmall{} organizes all widgets in the state according to a spatial order from top-left to bottom-right. This arrangement aligns with the natural browsing habits of users and forms a clear flow for state description. The ``Input Object'' portion in \figref{fig:functional-combine} (a) provides a concrete example of the state description for the ``S3" state in \figref{fig:illustrated-example}.

Ultimately, the state screenshot and the natural language description of the GUI hierarchy serve as multi-modal inputs. These are transmitted together to the Decision-Generation Agent to provide comprehensive environmental context for verifying specific functionalities.

\textbf{Action Execution.} Based on the test instructions output by the Decision-Generation Agent (see \secref{sec:strategy}), such as performing a click on a specific widget, this component is responsible for translating abstract test instructions into concrete actions. By invoking the underlying APIs of the mobile testing framework, this agent precisely executes predefined operations including clicks, text inputs, and swipes. Simultaneously, it monitors real-time state changes within the mobile application to facilitate assertion verification.

\textbf{Case Synthesis.} Upon the completion of the test sequence execution, this component performs a structured integration of the interaction events and corresponding assertions generated throughout the exploration process. By organizing these elements according to their execution order, this agent ultimately generates executable functional test cases including events and assertions. This process completes the transformation from dynamic interaction behaviors into standardized test cases for the target functionalities.

\subsubsection{Decision-Generation Agent}
\label{sec:strategy}

The Decision-Generation Agent serves as the central coordination module of \toolNameSmall{}, bearing the critical responsibility of formulating testing strategies. This agent performs dynamic decision-making based on three inputs, which are the current state information, the requirement description of the target functionality, and the core business logic derived from the first stage. These decision strategies subsequently drive the Perception-Interaction Agent to execute specific actions.

In the decision-making process, this agent faces \textbf{\emph{three challenges}}. First, business logic often originates from a generalized synthesis of multiple applications. For instance, removing an item might require a ``selection and click" in one app but a ``swipe gesture" in another. Consequently, only a subset of these steps may apply to the target mobile application, necessitating precise step selection and adaptation. Second, after identifying the applicable steps, the agent must accurately map them to specific GUI widgets within the current state. This mapping is essential to generate executable instruction descriptions for subsequent events and assertions. Finally, due to the inherent limitations and hallucination issues of LLMs regarding task termination~\cite{ji2023towards}, models often struggle to perceive task completion independently, leading to redundant exploration. Therefore, an effective completion determination mechanism is required to avoid generating unnecessary operations.

\begin{figure*}[t]
	\center
 \includegraphics[scale = 0.42]{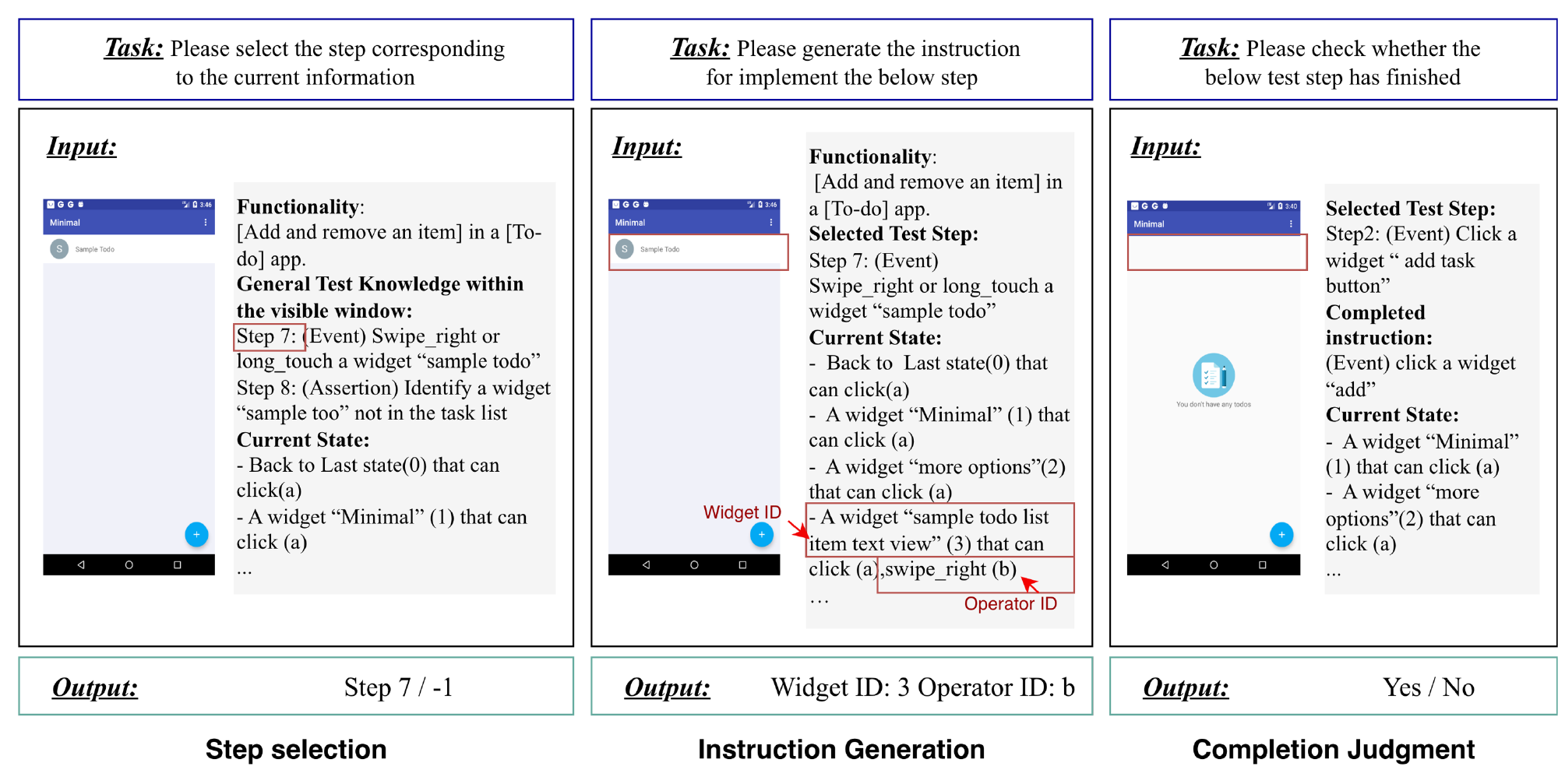}
	\caption{Example of Decision-Generation Agent.}
	\label{fig:functional-combine}
\end{figure*}

To address the preceding challenges, the Decision-Generation Agent incorporates three collaborative components comprising step selection, instruction generation, and completion judgment, as illustrated in \figref{fig:functional-combine}. The detailed logic of this process is further elaborated in Algorithm \ref{alg:Decision-Generation}.

\begin{algorithm}[h]
\caption{Decision-Generation Agent Execution Flow}
\label{alg:Decision-Generation}
\begin{algorithmic}[1]
\newcommand{\FUNCTION}[2]{\item[] \textbf{Function} \textsc{#1}(#2)}
\newcommand{\ENDFUNCTION}{\item[] \textbf{End Function}}
\newcommand{\CALL}[2]{\textsc{#1}(#2)}
\renewcommand{\algorithmiccomment}[1]{\hfill $\triangleright$ \textit{#1}}
\renewcommand{\algorithmicrequire}{\textbf{Input:}}
\renewcommand{\algorithmicensure}{\textbf{Output:}}

\REQUIRE Requirement Description ($R$), Core Business Logic ($L$), Current State ($S_{curr}$)
\ENSURE Test Case ($T$)

\STATE Initialize sliding window $W$ with size $Step_{num}$
\STATE $current\_step \leftarrow 0$; $is\_completed \leftarrow$ ``Yes''
\STATE $executed\_cmds \leftarrow \emptyset$; $Candidate\_Steps \leftarrow$ GetStepsInWindow($W, L$)
\WHILE{Task Not Complete}
    \STATE \COMMENT{ Step Selection}
    \IF{$is\_completed =$ ``Yes''}
        \STATE $selected\_step \leftarrow$ StepSelection($R, S_{curr}, Candidate\_Steps$)

        \IF{$selected\_step = (-1)$}
            \STATE \COMMENT{No applicable step found}
            \STATE Slide $W$ forward
            \STATE \textbf{continue}
        \ELSE
            \STATE $current\_step \leftarrow selected\_step$
            \STATE $is\_completed \leftarrow$ ``No''
        \ENDIF
    \ENDIF

    \STATE \COMMENT{ Instruction Generation}
    \STATE \CALL{GenerateInstruction}{$R, S_{curr}, current\_step$}

    \STATE \COMMENT{ Completion Judgment}
    \STATE $is\_completed \leftarrow$ CompletionJudgment($R, current\_step, S_{curr}$)

    \IF{$is\_completed =$ ``Yes''}
        \STATE $current\_step \leftarrow current\_step + 1$
        \STATE Reset attempts
    \ELSE
        \STATE Increment attempts
        \IF{attempts $>$ threshold}
            \STATE \COMMENT{Prevent infinite loops}
             \STATE Skip step and Reset attempts
        \ENDIF
    \ENDIF

    \STATE Update $S_{curr}$ \COMMENT{Perceive new state via PIA}
\ENDWHILE

\end{algorithmic}
\end{algorithm}

Algorithm ~\ref{alg:Decision-Generation} illustrates the detailed execution flow of the \textbf{Decision-Generation Agent}. First, the \textbf{step selection} component initializes a sliding window mechanism to focus on a limited set of candidate steps and identifies the specific logic step that best aligns with the current GUI state (Line 6). Second, the \textbf{instruction generation} component synthesizes concrete operation instructions for the selected step (Line 17), employing a differentiated processing mechanism to handle event-based interactions or assertion-based verifications. Third, the \textbf{completion judgment} component evaluates whether the current step is completed (Line 19) to decide whether to proceed to the next step or retry, while incorporating a threshold mechanism to prevent infinite loops (Lines 20-27).

\textbf{Step selection.} In this component, as illustrated in \figref{fig:functional-combine} (a), \toolNameSmall{} provides the LLM with three primary inputs: the requirement description of the target functionality, the core business logic generated by the Knowledge-Fusion Agent, and the GUI state information captured by the Perception-Interaction Agent.

To improve decision accuracy, \toolNameSmall{}  employs a sliding window mechanism to manage the sequence of logic steps through three primary actions (line 7). First, the system uses a window of size $Step_{num}$ to restrict the focus of the LLM to a specific range of candidate logic steps. This design allows the model to precisely evaluate the applicability of each step relative to the current GUI state by analyzing a limited set of possibilities. Second, the model performs reasoning according to the order of steps within the window to select the logic step that best matches the current context of the mobile application. Third, if no applicable step exists within the window, the model returns a specific identifier ``(-1)", which triggers the sliding window to update and move to a new set of candidate steps (lines 8-9). This approach ensures that the agent effectively adapts business logic to the specific functionalities of the target  application.

\textbf{Instruction Generation.} The instruction generation component, as illustrated in \figref{fig:functional-combine} (b), processes three inputs: the target requirement description, the specific logic step identified by the step selection component, and the current state information provided by the Perception-Interaction Agent (see Section \ref{sec:interaction}). Based on these inputs, the LLM generates concrete operation instructions, such as clicking the ``Add" button in state S1 of \figref{fig:illustrated-example}, and transmits them to the Perception-Interaction Agent for execution (line 17). To prevent redundant operations, \toolNameSmall{} maintains a record of completed test steps and their corresponding instructions. \toolNameSmall{} utilizes a differentiated processing mechanism for two distinct types of test steps.

\underline{Event-based steps.} An event includes a widget, an action, and an optional input. For event-based steps, the LLM returns a specific combination of widget ID and operation ID, as shown in \figref{fig:functional-combine} (b). This information is passed to the Perception-Interaction Agent and ultimately integrated into the final test case.

\underline{Assertion-based steps.} An assertion includes a widget and a condition. GUI testing primarily involves two types of conditions~\cite{lin2019test,zhao2020fruiter}. The first type of assertions verifies the existence of a widget on the current state, such as assertion A1 in \figref{fig:illustrated-example}. For this type, \toolNameSmall{} inputs the current state and the logic step into the LLM to retrieve the corresponding widget ID. The second type of assertions verifies the disappearance of a widget that appeared in a previous state, such as assertion A2 in \figref{fig:illustrated-example}. Since the target widget is absent from the current state, \toolNameSmall{} utilizes a state backtracking mechanism to identify the widget from historical states and retrieve its widget ID.
Finally, once the widget is identified based on the preceding process, \toolNameSmall{} generates the appropriate assertion based on the widget ID and the corresponding conditions, and sends this assertion to the Perception-Interaction Agent for verification.

Through this processing mechanism, \toolNameSmall{} effectively manages diverse logic step requirements. This ensures both the accuracy and executability of events and assertions for complex functionalities within the mobile application.

\textbf{Completion Judgment.} To address the inherent limitations and hallucination issues of LLMs, \toolNameSmall{} implements an active query mechanism, as illustrated in \figref{fig:functional-combine} (c). This mechanism decomposes the overall testing task into an ordered sequence of test steps and performs real-time state evaluation during the execution of each step.

Specifically, \toolNameSmall{} initiates the completion judgment process immediately after the LLM generates an operation instruction. During this phase, the model performs reasoning based on three features, which are the description of the current logic step, the sequence of already executed instructions, and the most recent state. The model must output a binary decision, where ``Yes" indicates step completion and ``No" indicates the step remains unfinished (line 19). If the step is judged as complete, \toolNameSmall{} proceeds to the selection phase for the next logic step (lines 20-21). If judged as incomplete, the agent continues to generate subsequent instructions targeting the current step (lines 24-26).

To prevent infinite loops during complex steps, \toolNameSmall{} enforces a maximum attempt limit. When the number of consecutive attempts for a single logic step reaches a predefined threshold, \toolNameSmall{} automatically skips that step and moves to the next one. This design ensures both thorough exploration of individual steps and the overall progress efficiency of the testing process. The active query mechanism offers two primary advantages. It maintains the integrity of the business logic through real-time state assessment and effectively handles complex scenarios requiring multiple operations, thereby ensuring high execution efficiency for the target mobile application and its functionalities.

\section{Evaluation}\label{sec:evaluation}

To evaluate \toolNameSmall{} comprehensively, we conduct the research questions focusing on five key perspectives. First, we assess the effectiveness of \toolNameSmall{} in generating functional test cases for mobile applications in real-world industrial scenarios and compare \toolNameSmall{} with baseline approaches. Second, we investigate the specific contributions of different techniques within \toolNameSmall{} through an ablation study. Third, we evaluate the robustness of \toolNameSmall{} by examining whether its performance remains consistent across various underlying foundation models.
Fourth, we analyze the efficiency of \toolNameSmall{} and the baselines.
Fifth, we analyze the generalizability of \toolNameSmall{} on new applications.
Details are as follows.

\textbf{RQ1:} How effective is \toolNameSmall{} compared with the baselines?

\textbf{RQ2:} How do \toolNameSmall{}'s main techniques affect the GUI test generation?

\textbf{RQ3:} How robust is \toolNameSmall{} across different foundation models?

\textbf{RQ4:} How efficient is \toolNameSmall{} compared with the baselines?

\textbf{RQ5:} How does \toolNameSmall{} perform on new applications?

\subsection{Experimental Setup}
\label{sec:experimental-setup}

\textbf{Experimental subjects.}  We utilize two widely-used functional testing benchmarks for evaluation, i.e.,  FrUITeR~\cite{fruiter-dataset} dataset and Lin~\cite{craftdroid-dataset} dataset. These benchmarks cover various industrial-grade applications, including ABC News~\cite{abc-news} and Firefox~\cite{firefox}, providing a solid foundation for evaluating \toolNameSmall{} in real-world scenarios. Both datasets provide the target applications and developer-written functional test cases. Specifically, the test cases in the Lin dataset contain complete event sequences and assertions, while the FrUITeR dataset provides only event sequences.

We collect all installable applications and executable test cases from both datasets to form our experimental subjects. To enhance data diversity, we invite two volunteer engineers with 3--5 years of Android development experience to independently write requirement descriptions based on the actual functionality of the test cases. This dual-annotation mechanism aims to capture the diversity in how different individuals describe the same functionality. The final experimental dataset comprises 28 applications, 95 test cases (serving as \emph{ground-truth test cases}), and 190 functional requirement descriptions (two independent descriptions per test case). Notably, compared with existing research~\cite{zhang2025appagent,wen2023empowering}, our evaluation involves the largest number of mobile applications and categories. Table~\ref{tab:experimental-subject} presents the detailed statistics of the experimental subjects.

\begin{table}[t]
  \centering
  \caption{Statistics of Experimental Subjects}
  \begin{tabular}{c|l|r|r|r|r|r}
    \toprule
    \textbf{Dataset} & \multicolumn{1}{c|}{\textbf{Category}} & \multicolumn{1}{c|}{\textbf{Apps}} & \multicolumn{1}{c|}{\textbf{Cases}} & \multicolumn{1}{c|}{\textbf{Events}} & \multicolumn{1}{c|}{\textbf{Assertions}} & \multicolumn{1}{c}{\textbf{Reqs}} \\
    \midrule
    \multirow{2}[2]{*}{FrUITeR} & News  & 4     & 32    & 76    & -     & 64 \\
          & Shopping & 3     & 21    & 116   & -     & 42 \\
    \midrule
    \multirow{5}[2]{*}{Lin} & Browser & 5     & 10    & 32    & 20    & 20 \\
          & To-Do & 5     & 10    & 39    & 15    & 20 \\
          & Shopping & 4     & 8     & 49    & 26    & 16 \\
          & Email & 2     & 4     & 14    & 12    & 8 \\
          & Calculator & 5     & 10    & 33    & 10    & 20 \\
    \midrule
    \textbf{Total} & -     & 28    & 95    & 359   & 83    & 190 \\
    \bottomrule
  \end{tabular}%
  \label{tab:experimental-subject}%
\end{table}%

\textbf{Baselines approaches.} To comprehensively evaluate the performance of \toolNameSmall{}, we select two state-of-the-art approaches from both academia and industry for comparison, i.e., AutoDroid~\cite{wen2023empowering} and AppAgent~\cite{zhang2025appagent}.

AutoDroid~\cite{wen2023empowering} adopts a technical route combining static semantic understanding with dynamic exploration. This approach first comprehends the application semantics through static analysis and subsequently guides LLMs to generate test cases based on requirement descriptions.

AppAgent~\cite{zhang2025appagent} is released by Tencent~\cite{tencent} Inc. (a fortune global 500 company). This approach learns to navigate and use new apps through two distinct modes: exploration and the observation of human demonstrations. By leveraging these learning mechanisms, it generates corresponding test cases for various functionalities.

Note that, existing functional test generation approaches generally suffer from a critical limitation involving the inability to generate effective assertions. As assertions represent core elements for verifying functional correctness, their absence severely restricts the practical value of these approaches in real-world industrial scenarios. In contrast, by incorporating retrieved domain knowledge to guide the LLM, \toolNameSmall{} is capable of generating functional test cases containing both complete event sequences and verification assertions for target applications.

\textbf{Evaluation metrics.}
To evaluate \toolNameSmall{} and the baselines, we follow related studies~\cite{zhang2026gui} and utilize two evaluation metrics, i.e., perfect-rate and success-rate.

\underline{Perfect-rate}: The proportion of generated test cases ($Test_{t}$) that are consistent with the ground-truth test cases ($Test_{gt}$). The evaluation process for perfect-rate is fully automated and requires no human intervention.

\begin{equation}
\textit{Perfect-rate} = \frac{\textit{Test}_{t} \cap \textit{Test}_{gt}}{\textit{Test}_{t}}
\end{equation}

\underline{Success-rate}: The proportion of generated test cases ($Test_{t}$) that successfully test the target functionality ($Test_{suc}$).

\begin{equation}
\textit{success-rate} = \textit{Test}_\textit{suc}\,/\,\textit{Test}_\textit{t}
\end{equation}

To quantify the generated test cases that successfully verify the target functionalities, we acknowledge that verification can be achieved through diverse sequences. For this reason, the developer-provided ground-truth represents only one feasible solution rather than an exhaustive standard. Since enumerating all potential valid cases is impossible, we systematically inspect those deviating from the ground-truth and consider them valid if they still fulfill the target requirement through semantically equivalent events and assertions. This manual verification involves the two volunteers previously responsible for requirement drafting. We provide them with an evaluation package containing the target application, the generated test cases, and the ground-truth test cases. To facilitate consistent judgment, we explicitly highlight the differences in events and assertions between each generated test case and its corresponding ground-truth test case. During evaluation, the volunteers execute both the ground-truth and generated test cases on the target application, and assess whether the generated test case still covers the essential functional steps and verification conditions required by the target functionality. For events and assertions, additional ones are considered acceptable if they do not hinder the testing objective, while replaced ones are regarded as valid if they produce the same functional effect as those in the ground-truth or correctly verify the expected functional outcome. Each volunteer first assesses validity independently, followed by group discussions to resolve any disagreements until reaching a final consensus. This multi-party mechanism maintains the reliability and objectivity of the evaluation results.

\textbf{Implementation details.}
Regarding text vectorization, \toolNameSmall{} utilizes the bge-base-en-v1.5~\cite{bge-base} model to convert requirement descriptions and functional summaries into embedding vectors. This model is widely applied in semantic similarity calculation tasks~\cite{li2025matching,gao2023retrieval,gao2023retrieval}.
Regarding parameter configuration, we optimize the three hyperparameters of \toolNameSmall{} through systematic preliminary experiments. For the sliding window size $Step_{num}$ and the number of retrieved similar test cases $Top_{sim}$, we conduct validation experiments on a candidate set \{1, 2, 3\} based on 20\% of the test cases. The results indicated that \toolNameSmall{} achieves optimal performance when $Step_{num}=2$ and $Top_{sim}=3$.
For the LLM temperature, AppAgent provides an official setting, and we therefore follow its original configuration with a temperature of 0.0. In contrast, AutoDroid does not specify an official temperature setting. Thus, for LogiDroid and AutoDroid, we tune the temperature using the candidate set \{0.0, 0.2, 0.4, 0.6, 0.8\}. The results show that both approaches achieve the best overall performance when the temperature is set to 0.4.
Therefore, we adopt this configuration for the final experiments.

\subsection{RQ1: Effectiveness}

\label{sec:rq1}

We evaluate \toolNameSmall{}'s effectiveness on the FrUITeR and Lin datasets, and compare it against two representative baselines: AutoDroid~\cite{wen2023empowering} and AppAgent~\cite{zhang2025appagent}. The performance is evaluated based on two metrics, i.e., perfect-rate and success-rate. For fairness, all approaches utilize GPT-5~\cite{singh2025openai} as the underlying LLM.

Notably, we implement a rigorous evaluation protocol to avoid data leakage in the retrieval process of \toolNameSmall{}. When generating test cases for the specific functionalities of a target application, \toolNameSmall{} \emph{removes all test cases associated with that application from the retrieval dataset}. Therefore, the system cannot access the ground-truth test case itself, nor any other test case from the same application. The retrieved test cases come from different applications with the target application, and thus involve different GUI implementations, interaction flows, and application contexts. In this sense, the retrieval source and evaluation target are separated at the application level, and the evaluation measures cross-application knowledge reuse rather than data leakage. By enforcing this data isolation, the evaluation provides an authentic reflection of \toolNameSmall{}'s functional test generation capabilities.

\begin{table}[t]

  \centering

  \caption{Overall effectiveness of \toolNameSmall{} and the baselines}

    \begin{tabular}{c|l|r|r}

    \toprule

    \textbf{Dataset} & \textbf{Approach} & \multicolumn{1}{l|}{\textbf{Perfect-rate}} & \multicolumn{1}{l}{\textbf{Success-rate}} \\

    \midrule

    \multirow{3}[6]{*}{FrUITeR} & \toolNameSmall{} & \textbf{20\%}  & \textbf{40\%} \\

\cmidrule{2-4}          & AppAgent & 9\%   & 32\% \\

\cmidrule{2-4}          & AutoDroid & 15\%  & 27\% \\

    \midrule

    \multirow{4}[8]{*}{Lin} & \toolNameSmall{} & 41\%  & 57\% \\

\cmidrule{2-4}          & \toolNameSmall{}* & \textbf{48\%}  & \textbf{65\%} \\

\cmidrule{2-4}          & AppAgent & 21\%  & 42\% \\

\cmidrule{2-4}          & AutoDroid & 10\%  & 27\% \\

    \bottomrule

    \end{tabular}%

    \vspace{4pt}

    \begin{minipage}{0.95\linewidth}
    \footnotesize
    \centering
    \toolNameSmall{}* denotes the variant evaluated without considering assertions.
    \end{minipage}
  \label{tab:effectiveness}%

\end{table}%

\textbf{Effectiveness results.} Table~\ref{tab:effectiveness} presents the evaluation results of \toolNameSmall{} compared with baseline approaches on the FrUITeR and Lin datasets. To ensure the reliability of the results, two volunteers (see \secref{sec:experimental-setup}) verify whether each generated test case successfully test the target functionality. We quantify inter-rate reliability using Fleiss' kappa coefficient~\cite{fleiss1971measuring}. The resulting value is 0.93, which satisfies the statistical standard of \emph{``almost perfect agreement"}. This metric indicates that the evaluation results possess high consistency and credibility.

On the FrUITeR dataset, \toolNameSmall{} demonstrates a significant advantage. Specifically, 40\% of the generated test cases successfully verified target functions, and 20\% are perfectly identical to the ground-truth. Compared to baseline approaches, this result represents an improvement of over 25\% in success-rate and over 33\% in perfect-rate.

\begin{figure*}[t]
	\center
 \includegraphics[width=0.8\linewidth]{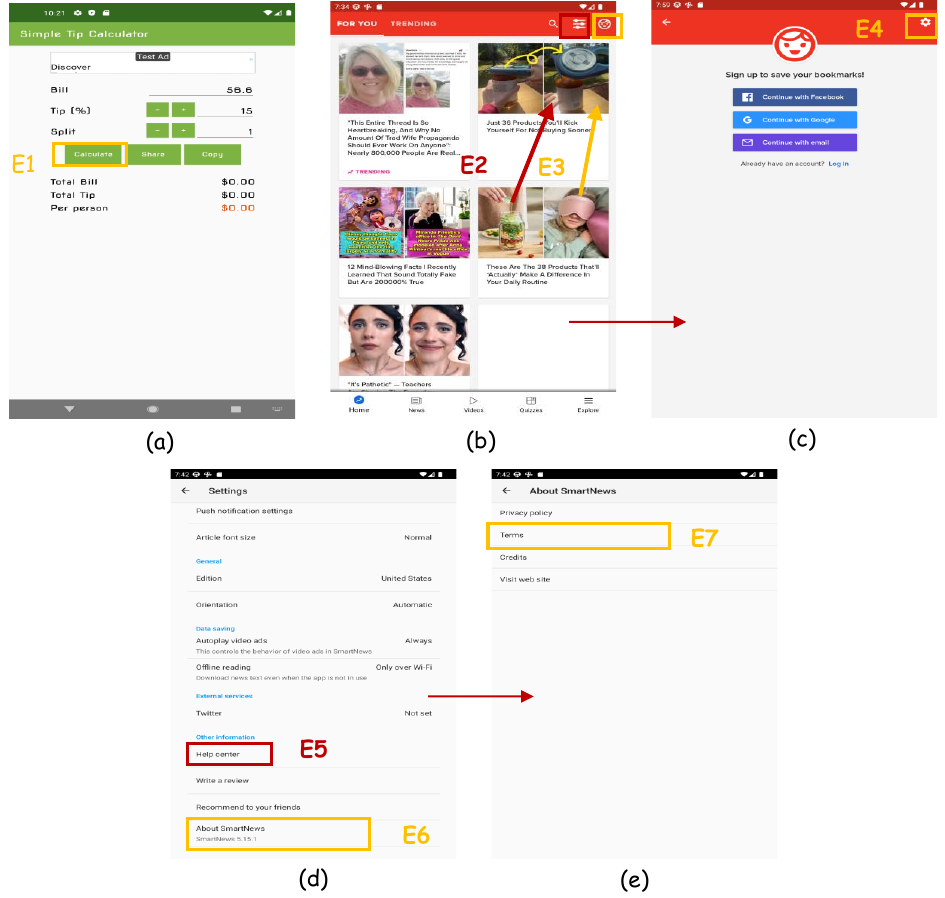}
	\caption{Illustrative Examples of Failure Cases}
	\label{fig:failcase}
\end{figure*}

On the Lin dataset, \toolNameSmall{} exhibits consistent performance improvements across tasks.
\toolNameSmall{} achieved a 57\% success-rate, with 41\% of test cases being perfectly identical to the ground-truth. Notably, the Lin dataset includes complete events and assertions, whereas baseline approaches only support event generation. Therefore, we additionally evaluated \toolNameSmall{}’s performance without considering assertions (denoted as \toolNameSmall{}*). In this setting, \toolNameSmall{}* achieved a 65\% success-rate and a 48\% perfect-rate. This outperforms baseline methods by over 55\% in success-rate and over 129\% in perfect-rate.

\textbf{Effectiveness analysis.} We observe three findings from Table~\ref{tab:effectiveness}.

First, compared to AppAgent, \toolNameSmall{} achieves a significant improvement in the perfect-rate, with increases of 122\% on the FrUITeR dataset and 129\% on the Lin dataset. This advantage primarily stems from \toolNameSmall{}’s knowledge retrieval and alignment mechanism. This mechanism extracts domain knowledge from historical test cases to generate high-quality business logic. In contrast, AppAgent lacks knowledge reuse capabilities. As a result, it generates test cases with numerous redundant operations, which compromise the overall quality of the test cases.

Second, compared to AutoDroid, \toolNameSmall{} achieves a significant improvement in the success-rate,
with increases of 48\% on the FrUITeR dataset and 141\% on the Lin dataset.
This performance boost is mainly attributed to \toolNameSmall{}'s context-aware test generation mechanism.
On one hand, it leverages multimodal information (i.e., textual and visual data) to deeply understand GUI semantics.
On the other hand, it decomposes complex tasks through a progressive decision-making process involving
step selection, instruction generation, and completion judgment.
In comparison, AutoDroid relies solely on textual information and employs a bulk generation strategy,
making it difficult to ensure the completeness of test cases.

Third, unlike existing approaches, \toolNameSmall{} can generate assertions. Assertions are crucial for functional testing. The domain knowledge refined during our Knowledge Retrieval and Fusion stage contains high-quality, reusable business logic. This enables \toolNameSmall{} to generate critical assertions required for effective functional verification. Conversely, AppAgent and AutoDroid only generate basic operation sequences, which limits their practical application value.

\textbf{Failure Analysis.} To understand the weaknesses of \toolNameSmall{}, we manually analyze all failed test cases and identify three main reasons. We select three examples (see Figure~\ref{fig:failcase}) to illustrate these failure reasons and also discuss possible ways to address them. In this figure, the yellow events represent the correct events and the red events represent the wrong events.

First, the extracted business logic may fail to cover all necessary test steps for the target functionality, which leads to incomplete generated test cases. For example, state (a) in Figure~\ref{fig:failcase} corresponds to the functionality of calculating the total amount for a bill of 56.6 with a 15\% tip. The intended test procedure should first complete the required input setup and then trigger the calculation. However, the extracted business logic only includes the calculation step (i.e., E1) while missing the input-setup step. As a result, the generated test case cannot fully test the target functionality. A potential way to address this problem is to check whether the extracted business logic covers all key steps of the target functionality after knowledge fusion.

Second, the instruction generation component may produce incorrect events for the given test steps, resulting in a generated test case with incorrect interactions. For example, states (b) and (c) of Figure~\ref{fig:failcase} correspond to the functionality of opening the contact page from application settings. The intended test procedure should open the profile entry, enter ``Settings'', and then open the contact entry. At the state (b), the instruction generation component produces the event of clicking the filter icon (i.e., E2), whereas the correct interaction should first be clicking the profile icon (i.e, E3) and then clicking the settings icon (i.e., E4). This happens because multiple candidate widgets in the current GUI state are semantically related, making the current generation process prone to confusion. Incorporating richer state information and widget attributes into the instruction generation component may help improve its effectiveness.

Third, the step selection component may choose an incorrect step in a hierarchical state, causing the generated test case to deviate from the intended functionality. We illustrate this problem using states (d) and (e) of Figure~\ref{fig:failcase}, which correspond to the functionality of opening the Terms page in an application. The intended test procedure should open ``Settings'', select the ``About'' entry, and then open ``Terms''. However, the step selection component does not select the ``About entry'' (i.e., E6) rather than randomly select E5. As a result, the generated test case fails to reach the target page and open ``Terms'' (i.e., E7). Incorporating hierarchical state information into the step selection process, and enabling it to better follow the intended procedure, may help reduce this problem.

\finding{\textbf{Answer to RQ1:} \toolNameSmall{} is effective and significantly outperforms the state-of-the-art baselines in functional testing scenarios. Furthermore, the generated test cases by \toolNameSmall{} includes assertions that the state-of-the-art baselines fail to.}

\subsection{RQ2: Main Techniques}

\label{sec:rq2}

To evaluate the impact of \toolNameSmall{}'s main techniques, we randomly select 50\% of the applications from each category in the Lin dataset as our research subjects. Note that, we choose the Lin dataset as it offers broader category and complete event sequences with assertions, making it more representative for analyzing the contribution of individual approaches.

\textbf{Experimental setup.} To assess the contribution of core techniques, we design a series of ablation experiments in a controlled setting. These experiments analyze the influence of different agents on overall system performance.

For the Knowledge Retrieval and Fusion stage, we conduct two sets of ablation experiments. First, we construct the ``\toolNameSmall{} (w/o Semantic-Retrieval Agent)'' variant. After receiving requirement descriptions, this variant skips the cross-application retrieval process. It generates business logic using only the Knowledge-Fusion Agent based on the LLM's internal knowledge. Second, we develop the ``\toolNameSmall{} (w/o Knowledge-Fusion Agent)'' variant. This variant passes the requirement descriptions and retrieval results directly to the subsequent stages without any fusion processing.

For the Context-Aware Test Generation stage, the Perceptual-Interaction Agent represents a fundamental component. Removing it prevents the system from obtaining GUI information and generating test cases. Thus, we only conduct an ablation experiment by removing the Decision-Generation Agent, creating the ``\toolNameSmall{} (w/o Decision-Generation Agent)'' variant. In this setting, the system skips the specialized decision-making strategy. Instead, it makes decisions via the LLM directly.

\begin{table}[t]
  \centering
  \caption{Effectiveness of \toolNameSmall{}'s Key Techniques}
    \begin{tabular}{l|c|c}
    \toprule
    \textbf{Approach}    & \textbf{Perfect-rate} & \textbf{Success-rate} \\
    \midrule
    \toolNameSmall{} & 30\%  & 70\% \\
    \midrule
    \toolNameSmall{} (w/o Semantic-Retrieval Agent) & 10\%  & 40\% \\
    \midrule
    \toolNameSmall{} (w/o Knowledge-Fusion Agent) & 5\%   & 20\% \\
    \midrule

    \toolNameSmall{} (w/o Decision-Generation Agent) & 10\%  & 30\% \\
    \midrule
    AppAgent & 20\% & 35\% \\
    \midrule
    AutoDroid & 0\% & 10\% \\
    \bottomrule
    \end{tabular}
  \label{tab:ablation}
\end{table}

\textbf{Results analysis.} Table~\ref{tab:ablation} presents the results of the ablation study. The first row establishes the baseline performance of the full \toolNameSmall{} on a 50\% sample of the Lin dataset. The subsequent three rows show the performance after removing each of the three agents. For reference, we also report the results of AppAgent and AutoDroid under the same experimental setting in this table.

The results demonstrate that the Semantic-Retrieval, Knowledge-Fusion, and Decision-Generation agents are all critical to the effectiveness of \toolNameSmall{}. Removing the Semantic-Retrieval Agent causes the success-rate to drop from 70\% to 40\%, which validates the vital role of case retrieval in domain knowledge extraction. Similarly, removing the Knowledge-Fusion Agent and the Decision-Generation Agent leads to success-rate declines to 20\% and 30\%, respectively. This proves that both agents are indispensable for distilling business logic and planning decisions.

Further analysis reveals three critical observations. First, the lack of a semantic-retrieval mechanism prevents \toolNameSmall{} from accessing external domain knowledge. This limitation forces the model  to rely solely on the internal parameterized knowledge, which compromises test quality. Second, the absence of the knowledge-fusion mechanism has an even more profound impact because \toolNameSmall{} cannot extract generalized business logic from related cases. Without this fusion, the generated test cases lack the flexibility to adapt to the diverse implementations of various functionalities. Third, removing the decision-generating mechanism deprives the system of its ability to perform incremental task decomposition, which reduces the precision of the testing strategy. These observations validate the architectural design of \toolNameSmall{} where each agent works in synergy to maintain the effective generation of functional test cases.

\finding{\textbf{Answer to RQ2:} \toolNameSmall{}'s techniques are essential and contribute substantially to its superior performance in functional test generation.}

\subsection{RQ3: Robustness Analysis}

\label{sec:rq3}

To evaluate the performance consistency of \toolNameSmall{} across different foundation models, we replace the underlying LLM from \emph{GPT-5}~\cite{singh2025openai} with \emph{Gemini-2.5}~\cite{comanici2025gemini} while keeping all other experimental settings constant. This experiment uses the same data source as RQ2, which includes 50\% of the applications in the Lin dataset. We compare \toolNameSmall{} against AppAgent~\cite{zhang2025appagent} and AutoDroid~\cite{wen2023empowering}. This setup ensures consistent experimental conditions for cross-model comparison and validates the architecture of \toolNameSmall{} to different foundation models.

\begin{table}[t]
  \centering
  \caption{Robustness Evaluation Results of \toolNameSmall{} and Baseline Approaches}
  \begin{tabular}{l|c|c|c}
    \toprule
    \textbf{Approach} & \textbf{LLM} & \textbf{Perfect-rate} & \textbf{Success-rate} \\
    \midrule
    \multirow{2}{*}{\toolNameSmall{}}
      & GPT-5      & 30\% & 70\% \\
\cmidrule{2-4}
      & Gemini-2.5 & 20\% & 60\% \\
    \midrule
    \multirow{2}{*}{\toolNameSmall{}*}
      & GPT-5      & 50\% & 75\% \\
\cmidrule{2-4}
      & Gemini-2.5 & 25\% & 75\% \\
    \midrule
    \multirow{2}{*}{AppAgent}
      & GPT-5      & 20\% & 35\% \\
\cmidrule{2-4}
      & Gemini-2.5 & 15\% & 25\% \\
    \midrule
    \multirow{2}{*}{AutoDroid}
      & GPT-5      & 0\% & 10\% \\
\cmidrule{2-4}
      & Gemini-2.5 & 10\% & 10\% \\
    \bottomrule
  \end{tabular}
      \vspace{4pt}

    \begin{minipage}{0.95\linewidth}
    \footnotesize
    \centering
\toolNameSmall{}* denotes the variant of \toolNameSmall{} evaluated without considering assertions.
    \end{minipage}
  \label{tab:robust}
\end{table}

\textbf{Results statistics.} Table~\ref{tab:robust} illustrates the performance of various approaches on a 50\% sample of the Lin dataset using Gemini-2.5 as the foundation model. The Lin dataset includes complete event sequences and assertions, which the baseline approaches AppAgent and AutoDroid cannot generate. To maintain a fair evaluation, we record the performance of \toolNameSmall{} without considering assertions, denoted as \toolNameSmall{}*. Under the full evaluation with assertions, \toolNameSmall{} achieved a 60\% success-rate and a 20\% perfect-rate. When assertions are excluded, \toolNameSmall{}* reached a 75\% success-rate and a 25\% perfect-rate. Compared to the baselines, \toolNameSmall{} improves the success-rate by over 67\%, and the perfect-rate by over 200\%.

\textbf{Results analysis.} The experimental results indicate that \toolNameSmall{} maintains excellent performance across different foundation models, demonstrating high model-agnosticism and robustness. Replacing the underlying LLM from GPT-5 with Gemini-2.5 does not compromise \toolNameSmall{}'s performance advantage over baseline approaches. This result illustrates that the core architecture of \toolNameSmall{} adapts well to various foundation models. This trait is particularly crucial for practical applications. Users can adapt foundation models to their resource constraints and requirements without compromising performance stability.

\finding{\textbf{Answer to RQ3:} \toolNameSmall{} achieves stable performance across diverse foundation models, demonstrating the robust generalization of its architecture.}

\subsection{RQ4: Efficiency}
\label{sec:rq4}

To evaluate the efficiency of \toolNameSmall{} and baseline approaches, we performed experiments on the Lin dataset. We measure the average runtime and token usage for generating a single test case using \toolNameSmall{}, AppAgent, and AutoDroid.

\begin{table}[t]
  \centering
  \caption{Efficiency Statistics of \toolNameSmall{} and Baseline Approaches}
    \begin{tabular}{l|c|c}
    \toprule
    Approach    & Runtime [minutes] & Token \\
    \midrule
    \toolNameSmall{} & 6     & 31,796\\
    \midrule
    AppAgent & 8     & 34,936 \\
    \midrule
    AutoDroid & 5     & 18,374 \\
    \bottomrule
    \end{tabular}
  \label{tab:efficiency}
\end{table}

\textbf{Efficiency result.} Table~\ref{tab:efficiency} presents the comparison of resource consumption during the test case generation process. The results show that \toolNameSmall{} takes an average of 6 minutes to generate a single test case, with an average token consumption of 31,796. Compared with AppAgent, \toolNameSmall{} reduces runtime by 25\% and token consumption by 9\%. Compared with AutoDroid, \toolNameSmall{} increases runtime by 20\%  and token consumption by 73\%.

\textbf{Efficiency analysis.} Regarding resource efficiency, the token consumption of \toolNameSmall{} is lower than that of AppAgent but higher than that of AutoDroid. In terms of time efficiency, \toolNameSmall{} similarly ranks between the two baseline approaches.

Deep analysis reveals that the relatively high token consumption of \toolNameSmall{} primarily stems from the progressive decision-making mechanism of Decision-Generation Agent. This mechanism maintains the precision of test steps through multiple rounds of reasoning. Although there is room for optimization in runtime efficiency, the current efficiency level remains acceptable considering the significant improvements in testing accuracy (as shown in Section \ref{sec:rq1}). Notably, through the effective knowledge reuse mechanism, \toolNameSmall{} sustains high-quality test generation while achieving superior token efficiency compared to AppAgent.

\finding{\textbf{Answer to RQ4:} \toolNameSmall{}'s efficiency is comparable to the baselines.}

\subsection{RQ5: Generalizability}

To further evaluate the generalizability of \toolNameSmall{}, we conduct an additional study on a new dataset (i.e., the TEM dataset~\cite{tem-dataset}). The LLM used in this evaluation is GPT-5.

\textbf{Experimental setup.} The TEM dataset~\cite{tem-dataset} consists of five newly collected popular applications from the Google Play Store~\cite{google-play} and ten corresponding functional test cases, covering five representative categories, i.e., browser, to-do, shopping, mail, and calculator, as summarized in Table~\ref{tab:rq5-apps}. Since all applications in this dataset are newly introduced, the dataset provides an opportunity to further evaluate whether \toolNameSmall{} remains effective on previously unseen applications. Note that the TEM dataset only provides the application APKs and the target functionalities to be tested, but does not include ground-truth test cases. Therefore, we invite the same two volunteers (see \secref{sec:experimental-setup}) to construct the ground-truth test cases based on the given target functionalities. To assess the effectiveness of \toolNameSmall{} on this dataset, we follow the same evaluation process as described in \secref{sec:experimental-setup} to evaluate \toolNameSmall{}, AppAgent, and AutoDroid.

\textbf{Results.} Table~\ref{tab:rq5-results} presents the results of \toolNameSmall{} and the baselines on the TEM dataset. \toolNameSmall{} achieves a perfect-rate of 40\% and a success-rate of 70\%, outperforming AppAgent (20\% and 50\%, respectively) and AutoDroid (10\% and 30\%, respectively).
These results show that the effectiveness of \toolNameSmall{} is not limited to the original evaluation benchmarks. Even on new applications, \toolNameSmall{} still achieves strong performance, further demonstrating its ability to reuse and adapt business logic across different applications.

\begin{table}[tbp]
  \centering
  \caption{The TEM dataset used in RQ5}
  \small
    \begin{tabular}{l|l|l|l|r|r|r}
    \toprule
    \multicolumn{1}{c|}{\textbf{App}} & \multicolumn{1}{c|}{\textbf{Version}} & \multicolumn{1}{c|}{\textbf{Size}} & \multicolumn{1}{c|}{\textbf{Download}} & \multicolumn{1}{c|}{\textbf{Case}} & \multicolumn{1}{c|}{\textbf{Event}} & \multicolumn{1}{c}{\textbf{Assertion}} \\
    \midrule
    Web Browser & 20.11.22 & 5M    & 5K+   & 2     & 6     & 4 \\
    Done  & 1.0 & 1.6M  & 50K+  & 2     & 5     & 3 \\
    FiveMiles & 8.4.0 & 25.4M & 10M+  & 2     & 10    & 2 \\
    Pro Mail & 14.64.0 & 98.2M & 1M+   & 2     & 7    & 3 \\
    Tip Calculator & 2.6.2 & 4.3M  & 5K+   & 2     & 6     & 2 \\
    \bottomrule
    \end{tabular}%
  \label{tab:rq5-apps}%
\end{table}%

\begin{table}[t]
  \centering
  \caption{Results on the TEM dataset}
  \label{tab:rq5-results}
  \small
  \begin{tabular}{l|c|c}
    \toprule
    \textbf{Approach} & \textbf{Perfect-rate} & \textbf{Success-rate} \\
    \midrule
    \toolNameSmall{} & 40\% & 70\% \\
    \midrule
    AppAgent         & 20\% & 50\% \\
    \midrule
    AutoDroid        & 10\% & 30\% \\
    \bottomrule
  \end{tabular}
\end{table}

\finding{
\textbf{Answer to RQ5:}  The results indicate that \toolNameSmall{} remains effective on previously unseen applications and achieves consistently better performance than the baselines on the new benchmark.
}

\subsection{Threat to validity} We discuss potential threats to validity of our study across three primary dimensions.

A possible threat to external validity is the generalizability of our findings to other datasets. To mitigate this threat, we utilize a substantial number of applications and categories, exceeding the scale of most related work. Furthermore, we adopt all popular datasets in the field of functional testing as evaluation benchmarks. These datasets include various complex industrial mobile application examples that represent real-world scenarios. The diversity of these applications helps verify that \toolNameSmall{} effectively handles a wide range of functionalities.

A possible threat to internal validity involves potential errors in our implementation and experiments. To mitigate this threat, we manually analyze the test cases that fail to test the target functionalities. We also publish our implementation and experimental data to facilitate external validation. Regarding human evaluation, we invite two experienced developers as volunteers and provide them with a clear evaluation process. The resulting Fleiss' kappa coefficient of 0.93 confirms high consistency among evaluators, which maintains the objectivity of the assessment.

A possible threat to construct validity is evaluation metrics. To mitigate this threat, we follow related studies by utilizing two primary metrics to validate the effectiveness of the generated test cases. These metrics offer a reliable and objective basis for assessing the quality of the test cases generated by both the baseline approaches and \toolNameSmall{}.

\section{Discussion}\label{sec:discussion}

\begin{table}[t]
  {
  \centering
  \caption{Method coverage of \toolNameSmall{} and AutoDroid}
  \label{tab:code-coverage}%
  \small
  \setlength{\tabcolsep}{4pt}
  \renewcommand{\arraystretch}{0.96}

  \begin{tabular}{c!{\vrule width 0.5pt}c!{\vrule width 0.5pt}c!{\vrule width 0.5pt}c!{\vrule width 0.5pt}c!{\vrule width 0.5pt}c!{\vrule width 0.5pt}c}
    \toprule
    \multirow[c]{2}{*}{\textbf{\insertedpart{C}\insertedpart{a}\insertedpart{t}\insertedpart{e}\insertedpart{g}\insertedpart{o}\insertedpart{r}\insertedpart{y}}} &
    \multirow[c]{2}{*}{\textbf{\insertedpart{A}\insertedpart{p}\insertedpart{p}}} &
    \multicolumn{3}{c!{\vrule width 0.5pt}}{\textbf{\insertedpart{C}\insertedpart{o}\insertedpart{v}\insertedpart{e}\insertedpart{r}\insertedpart{a}\insertedpart{g}\insertedpart{e} \insertedpart{n}\insertedpart{u}\insertedpart{m}\insertedpart{b}\insertedpart{e}\insertedpart{r}}} &
    \multicolumn{2}{c}{\textbf{\insertedpart{C}\insertedpart{o}\insertedpart{v}\insertedpart{e}\insertedpart{r}\insertedpart{a}\insertedpart{g}\insertedpart{e} \insertedpart{c}\insertedpart{a}\insertedpart{p}\insertedpart{a}\insertedpart{b}\insertedpart{i}\insertedpart{l}\insertedpart{i}\insertedpart{t}\insertedpart{y}}} \\
    \cmidrule(lr){3-5} \cmidrule(lr){6-7}
    & &
    \textbf{\insertedpart{G}\insertedpart{T}\insertedpart{.}} &
    \textbf{\insertedpart{L}\insertedpart{o}\insertedpart{g}\insertedpart{i}\insertedpart{.}} &
    \textbf{\insertedpart{A}\insertedpart{u}\insertedpart{t}\insertedpart{o}\insertedpart{.}} &
    \textbf{\insertedpart{L}\insertedpart{o}\insertedpart{g}\insertedpart{i}\insertedpart{.}} &
    \textbf{\insertedpart{A}\insertedpart{u}\insertedpart{t}\insertedpart{o}\insertedpart{.}} \\
    \midrule

    \multirow{3}{*}{\insertedpart{Browser}}
    & \insertedpart{Privacy Browser} & \insertedpart{1566} & \insertedpart{1693} & \insertedpart{1559} & \insertedpart{99.7\%}  & \insertedpart{99.4\%} \\
    & \insertedpart{FOSS Browser}    & \insertedpart{739}  & \insertedpart{749}  & \insertedpart{677}  & \insertedpart{99.6\%} & \insertedpart{91.6\%} \\
    & \insertedpart{Firefox}         & \insertedpart{3068} & \insertedpart{3078} & \insertedpart{3290} & \insertedpart{100.0\%} & \insertedpart{99.3\%} \\
    \midrule

    \multirow{5}{*}{\insertedpart{To-Do}}
    & \insertedpart{Minimal}       & \insertedpart{2083} & \insertedpart{2075} & \insertedpart{1478} & \insertedpart{99.5\%} & \insertedpart{63.5\%} \\
    & \insertedpart{Clear List}    & \insertedpart{1770} & \insertedpart{1497} & \insertedpart{1520} & \insertedpart{84.6\%}  & \insertedpart{80.0\%} \\
    & \insertedpart{To-Do}         & \insertedpart{1874} & \insertedpart{1787} & \insertedpart{1649} & \insertedpart{95.4\%}  & \insertedpart{87.2\%} \\
    & \insertedpart{Simply Do}     & \insertedpart{91}   & \insertedpart{80}   & \insertedpart{80}   & \insertedpart{87.9\%}  & \insertedpart{87.9\%} \\
    & \insertedpart{Shopping List} & \insertedpart{1756} & \insertedpart{2196} & \insertedpart{1546} & \insertedpart{91.2\%}  & \insertedpart{78.3\%} \\
    \midrule

    \multirow{4}{*}{\insertedpart{Shopping}}
    & \insertedpart{Geek}  & \insertedpart{5757}  & \insertedpart{5624}  & \insertedpart{5153}  & \insertedpart{97.3\%} & \insertedpart{88.7\%} \\
    & \insertedpart{Yelp}  & \insertedpart{15212} & \insertedpart{15392} & \insertedpart{15220} & \insertedpart{99.2\%} & \insertedpart{98.9\%} \\
    & \insertedpart{Etsy}  & \insertedpart{9109}  & \insertedpart{8568}  & \insertedpart{7776}  & \insertedpart{91.0\%} & \insertedpart{85.4\%} \\
    & \insertedpart{Wish}  & \insertedpart{6962}  & \insertedpart{6938}  & \insertedpart{6602}  & \insertedpart{99.4\%} & \insertedpart{93.7\%} \\
    \midrule

    \multirow{2}{*}{\insertedpart{Mail}}
    & \insertedpart{K-9 Mail}   & \insertedpart{3065} & \insertedpart{3052} & \insertedpart{1153} & \insertedpart{99.5\%} & \insertedpart{37.6\%} \\
    & \insertedpart{Fast Email} & \insertedpart{2150} & \insertedpart{2150} & \insertedpart{1419} & \insertedpart{99.9\%} & \insertedpart{65.7\%} \\
    \midrule

    \multirow{4}{*}{\insertedpart{Calculator}}
    & \insertedpart{Tip Calculator} & \insertedpart{123}  & \insertedpart{118}  & \insertedpart{118}  & \insertedpart{95.9\%}  & \insertedpart{95.9\%} \\
    & \insertedpart{Simple Tip}     & \insertedpart{1124} & \insertedpart{1107} & \insertedpart{1103} & \insertedpart{98.5\%}  & \insertedpart{97.9\%} \\
    & \insertedpart{Tip Plus}       & \insertedpart{1529} & \insertedpart{1537} & \insertedpart{1515} & \insertedpart{100.0\%} & \insertedpart{99.1\%} \\
    & \insertedpart{Free Tip}       & \insertedpart{951}  & \insertedpart{942}  & \insertedpart{934}  & \insertedpart{99.1\%}  & \insertedpart{98.2\%} \\
    \midrule

    \multicolumn{2}{c!{\vrule width 0.5pt}}{\textbf{\insertedpart{Overall Average}}} &
    \textbf{\insertedpart{3274}} & \textbf{\insertedpart{3255}} & \textbf{\insertedpart{2933}} &
    \textbf{\insertedpart{96.5\%}} & \textbf{\insertedpart{86.0\%}} \\
    \bottomrule
  \end{tabular}
  }

\end{table}

\textbf{Test case collection.} The key innovation of \toolNameSmall{} lies in extracting and fusing domain-specific knowledge into  application functional testing. This process requires the systematic collection of functional test cases in real-world environments. The collection is feasible in practice, primarily based on the following two conditions.

First, a vast number of similar applications and their corresponding GUI test case resources exist in real-world environments~\cite{hu2018appflow,mariani2021evolutionary,zhao2020fruiter}. Mainstream platforms like Google Play~\cite{google-play} and F-Droid~\cite{fdroid} typically adopt standardized classification systems to group applications into specific categories such as Shopping and News. Meanwhile, open-source platforms like GitHub contain numerous mobile application projects with complete test cases, providing a rich source for the systematic collection of diverse functionalities.

Second, we develop a specialized automated collection tool that periodically crawls test cases from various channels and automatically generates functional summaries based on the method described in Section \ref{sec:search}. The entire process operates without human intervention, which provides a reliable technical foundation for expansion and maintenance of large-scale test repositories. This automated workflow maintains the scalability of our constructed dataset while verifying that new entries align with existing knowledge structures.

\textbf{Method coverage of LogiDroid.} Method coverage is a commonly-used code coverage metric~\cite{wang2018empirical, hu2024enhancing, imran2024empirical}. It shows whether the generated test cases can execute the methods related to the target functionality. In \secref{sec:rq1}, we evaluate LogiDroid using success-rate and perfect-rate.  To provide a more comprehensive understanding of the quality of the test cases generated by LogiDroid, we further evaluate their method coverage and compare it with that of the ground-truth test cases. For reference, we also evaluate the method coverage achieved by AutoDroid. Specifically, we design a new metric, \emph{Coverage-capability}. The coverage-capability is calculated as the ratio of common covered methods ($Cover_{common}$) between the generated test case and the ground-truth test case, to the total covered methods ($Cover_{gt}$) by the ground-truth test case:

\begin{equation}
Coverage\text{-}capability = \frac{Cover_{common}}{Cover_{gt}}
\end{equation}

We use WALLMauer~\cite{auer2024wallmauer} to instrument the target apps and calculate the method coverage for each app. Note that, coverage-capability and the metrics in \secref{sec:rq1} are evaluated from different perspectives. In this way, the result trends of LogiDroid and AutoDroid presented here and in \secref{sec:rq1} may be related but different.

\tabref{tab:code-coverage} shows the method coverage results for LogiDroid (denoted as Logi.) and AutoDroid (denoted as Auto.) for 18 applications that can be instrumented by WALLMauer~\cite{auer2024wallmauer} on the Lin dataset. For each approach, we report the number of covered methods and the coverage-capability. We also report the number of methods covered by the ground-truth test cases (denoted as GT.)  for reference. On average, LogiDroid reaches 96.5\% of the method coverage achieved by the ground-truth test cases, whereas AutoDroid reaches only 86.0\%. In terms of the number of covered methods, LogiDroid achieves an average of 3255, which is closer to the ground-truth value of 3274 than AutoDroid's 2933. These results indicate that LogiDroid achieves the closest match to the ground-truth in terms of method coverage, demonstrating satisfied capability in generating high-quality test cases that effectively exercise the implementation logic of the target functionality.

\textbf{Hallucination mitigation for semantic inconsistencies.}
While the current hallucination mitigation mechanism effectively maintains structural correctness and rule compliance, an important complementary aspect is the consideration of semantic inconsistencies. In practice, even when generated outputs satisfy predefined formatting constraints, they may still exhibit deviations from the intended functionality, such as omitting critical interaction steps, and generating assertions that do not faithfully reflect the expected outcomes. These inconsistencies arise from the challenge of maintaining stable alignment between abstract business logic and concrete application contexts, especially in dynamic and multimodal environments.

To further enhance reliability, three strategies may address semantic inconsistencies. First, a semantics-aware validation mechanism could be introduced by representing the extracted business logic as structured semantic constraints and then checking whether each generated test action remains consistent with these constraints. Second, an assertion-level checking mechanism could be incorporated by deriving expected functional outcomes from the extracted business logic and comparing them against the generated assertions, so as to assess whether generated assertions faithfully reflect the expected functional outcomes. Third, a cross-step consistency analysis mechanism could be developed to capture subtle semantic deviations through dependency analysis and temporal consistency constraints in long-horizon test generation. Exploring these strategies would complement the existing mitigation mechanism and further strengthen the correctness and robustness of generated test cases.

\textbf{Scalability in large-scale knowledge repositories.}
Although the current evaluation is conducted on a dataset containing 294 functional test cases, this scale is sufficient for assessing the effectiveness of LogiDroid. Looking toward real-world deployment with substantially larger repositories, the scalability of LogiDroid remains feasible for two main reasons. First, functional test repositories in real-world settings are typically not flat collections, but are naturally organized by application category, functionality type, and business scenario, which provides an initial structure for narrowing the retrieval space. Second, LogiDroid only requires a small set of highly relevant candidates for subsequent knowledge fusion, rather than processing the entire repository during each generation process. Therefore, the framework does not rely on exhaustively traversing over the full repository for each target functionality, which helps keep the retrieval-and-generation pipeline scalable.

\textbf{Effect of semantic summaries on retrieval accuracy.}
LogiDroid relies on condensed semantic summaries to support knowledge retrieval, which raises the concern that such summaries may omit important functional details and thereby reduce retrieval accuracy in large repositories. This risk is mitigated in LogiDroid for two main reasons. First, the semantic summaries are not produced through naive compression. Instead, they are generated through a structured summarization process over functional test cases, where the model is guided to focus on the core functional intent while taking as input semi-structured events, assertions, application category, and key widget attributes. This design helps preserve the functionality-level semantics needed for retrieval, rather than retaining only superficial textual patterns. Second, LogiDroid does not rely on a single summary for direct knowledge reuse. Instead, it combines category-based filtering with vector-based semantic matching to retrieve a small set of highly relevant candidates, and then performs knowledge fusion over multiple retrieved test cases to extract reusable business logic. As a result, even if an individual summary misses certain details, the subsequent multi-case retrieval and fusion process still provides a robust basis for accurate knowledge identification in large-scale repositories.

\begin{table}[t]
  \centering
  \caption{LogiDroid Results based on the number of similar test cases}
    \begin{tabular}{r|r|r|r}
    \toprule
    \multicolumn{1}{c|}{\textbf{Similar Test}} & \multicolumn{1}{c|}{\textbf{Test Num}} & \multicolumn{1}{c|}{\textbf{Avg. Perf.}} & \multicolumn{1}{c}{\textbf{Avg. Suc.}} \\
    \midrule
    0     & 3     & 0\%   & 17\% \\
    \midrule
    1     & 4     & 50\%  & 75\% \\
    \midrule
    2     & 3     & 33\%  & 33\% \\
    \midrule
    3     & 16    & 22\%  & 44\% \\
    \midrule
    $\geq$ 4   & 69    & 30\%  & 49\% \\
    \bottomrule
    \end{tabular}%
  \label{tab:similar-test-case}%
\end{table}%

\textbf{Influence of the number of similar test cases on LogiDroid.} We conduct a statistical analysis to evaluate how the number of similar historical test cases retrieved for a target functionality affects the effectiveness of LogiDroid. \tabref{tab:similar-test-case} shows the average perfect-rate (denoted as Avg. Perf.) and success-rate (denoted as Avg. Suc.) of LogiDroid with different numbers of similar test cases in the combined Lin and FrUITeR datasets. The results show that, as the number of similar test cases increases, the effectiveness of LogiDroid generally improves. For example, the success-rate increases from 17\% when no similar test case is available to 49\% when at least four similar test cases are available. This result supports the intuition of LogiDroid that extracting business logic from multiple similar test cases is beneficial for functional test generation.

Note that, when the number of similar test cases is one, the success-rate is relatively high (i.e., 75\%). However, this result is based on only four cases, which makes the statistics less stable. In addition, these cases are mainly from email applications, where the target applications and the retrieved source applications (i.e., the applications producing similar test cases) are highly alike. When the number of similar test cases is zero, LogiDroid still achieves a success-rate of 17\%. This result indicates that LogiDroid still retains some capability to generate useful test cases from requirement descriptions and GUI reasoning even without similar historical cases.

\section{Related Work}\label{sec:related}

\textbf{Functional test generation.}
Existing research proposes various functional test generation approaches for different operating systems. Regarding the Android system, Li et al.~\cite{li2020mapping} proposed a test generation approach based on a matching model. This approach relies on manually written test logics and manually filtered application screenshots, selecting operation events appearing in the interface through model matching. DroidBot-GPT~\cite{wen2023droidbot} first introduces LLMs to select operation events based on manually written test logics. AutoDroid~\cite{wen2023empowering}, as an enhanced version of DroidBot-GPT, introduces an offline state relationship understanding mechanism to improve testing effectiveness. LLMDroid~\cite{wang2025llmdroid} leverages LLM guidance to strategically direct testing towards unexplored functionalities, thereby enhancing automated exploration coverage. AppAgent~\cite{zhang2025appagent} and QTypist~\cite{liu2023fill} learn the functional operation logic of applications and generates corresponding test cases through two modes, which are autonomous exploration and observing human demonstrations.Similarly, ReuseDroid~\cite{li2025reusedroid}, LLMigrate~\cite{beyzaei2025automated} and Guardian~\cite{ran2024guardian} leverage LLMs to extract test intentions or enforce runtime constraints, generating functional test cases for target applications through a dynamic reasoning mechanism. Regarding the iOS system, AXNav~\cite{taeb2023axnav} and ILvuUI~\cite{jiang2023iluvui} both use test logics and application screenshots as inputs, utilizing vision-based LLMs for event selection. Regarding the windows system, AssistGUI~\cite{gao2023assistgui} is a GUI test generation framework specifically designed to adapt to that system.

There are two key differences between existing approaches with \toolNameSmall{}. First, for knowledge utilization, existing approaches mainly rely on the automatic exploration of the target application or the knowledge from the general LLMs, which affects the accuracy of testing.  On the contrary, we propose a retrieval-augmented architecture to synthesize testing expertise from cross-application historical data in a principled manner. This strategy transforms the test generation process from undirected exploration into a knowledge-driven reasoning task, enabling \toolNameSmall{} to navigate complex functionalities of the mobile application with higher precision.
Second, assertions are crucial for functional testing, yet existing approaches can only generate events and are unable to generate assertions, which limits their practical application value. However, \toolNameSmall{} supports the concurrent generation of both event sequences and semantic assertions.

\textbf{Bug detection for mobile applications.}
Based on different exploration strategies, bug detection approaches for mobile applications can be classified into four categories, which are random testing~\cite{monkey,machiry2013dynodroid, sun2023characterizing, sun2023property}, model-based approaches~\cite{su2017guided,baek2016automated,gu2017aimdroid,lai2019goal, li2017droidbot, su2021fully, wang2022detecting, liu2022guided, yu2024practical,sun2024property}, systematic testing approaches~\cite{anand2012automated,gao2018AndroidTest,mao2016sapienz}, and learning-based approaches~\cite{spieker2017reinforcement,borges2018guiding,koroglu2018qbe,li2019humanoid, liu2022nighthawk, yu2024practical-intrusive, qian2020roscript, bernal2020translating}. Representative research include Monkey~\cite{monkey} that adopts random exploration, AIMDROID~\cite{gu2017aimdroid} and Stoat~\cite{su2017guided} that combine static and dynamic analysis, SCENTEST~\cite{yu2024practical} that utilizes event knowledge graphs to guide the exploration process, and SynthesiSE~\cite{gao2018AndroidTest} that can dynamically infer Android model expressions.

The fundamental distinction between previous research and \toolNameSmall{} lies in different purposes. Existing approaches primarily focus on general GUI exploration and bug detection for the mobile application, yet they struggle to generate test cases that correspond to the specific requirements of individual functionalities. In contrast, \toolNameSmall{} shifts from undirected exploration to a knowledge-driven paradigm that generates precise functional test cases for diverse requirements.

\textbf{LLM-based code generation.}
General Large Language Models like ChatGPT~\cite{achiam2023gpt,singh2025openai} demonstrate significant potential in software engineering tasks, particularly in code generation, as evidenced by empirical study~\cite{huang2023empirical}. This success promotes the development of specialized models such as AlphaCode~\cite{li2022competition}, CodeGen~\cite{nijkamp2022codegen}, and Incoder~\cite{fried2022incoder}. Researchers obtain these models by fine-tuning general LLMs with code corpora or through specialized training. Furthermore, a series of research studies have conducted in-depth exploration in the decoding stage. Zhang et al.~\cite{zhang2023planning} propose a planning-guided decoding algorithm based on Monte Carlo Tree Search, which improves code quality by exploring diverse program generation paths. Shi et al.~\cite{shi2022natural} optimize output results by generating multiple program samples combined with test case re-ranking. These approaches fully exploit the potential of LLMs in program generation and debugging.

The key distinction between \toolNameSmall{} and existing LLMs for code generation lies in their object characters. Existing LLMs in code generation primarily focus on converting general natural language descriptions into general program code.  They lack the capability to manage the complex interfaces and dynamic GUI states of a mobile application. In contrast, \toolNameSmall{} dynamically explores the target mobile application to capture real-time widget information and GUI state transitions. By learning domain-specific testing logic from these interactions, \toolNameSmall{} guides the generation process to emulate real-world user behavior for diverse functionalities.

\section{Conclusion}\label{sec:conclusion}

In this paper, we propose \toolNameSmall{}, a two-stage approach that addresses the challenges of automated functional test case generation for mobile applications. By constructing a two-stage architecture, \toolNameSmall{} effectively resolves the core issues of domain knowledge acquisition and functional semantic understanding. Specifically, the approach utilizes Semantic-Retrieval and Knowledge-Fusion agents to achieve systematic accumulation and reuse of test knowledge, while it leverages Perception-Interaction and Decision-Generation agents to enable adaptive and progressive test generation in dynamic environments.
Our evaluation of \toolNameSmall{} on 28 mobile applications and 190 functional requirements demonstrates that the approach significantly outperforms the state-of-the-art approaches, achieving substantial improvements in success-rate and perfect-rate. Notably, \toolNameSmall{} also exhibits advantages in generating test cases with complete verification assertions for diverse functionalities, providing a practical solution for large-scale industrial deployment.

In the future, we plan to focus on two aspects. First, we aim to optimize the decision-making mechanism under dynamic states, investigating the integration of reinforcement learning techniques to establish smarter exploration strategies. Second, we plan to construct a cross-platform test generation paradigm by establishing a unified abstraction approach, enabling the intelligent migration and adaptation of test cases across different mobile platforms.


\bibliographystyle{ACM-Reference-Format}
\bibliography{main}

\end{document}